\documentclass[%
 aip,
 jcp,
 amsmath,amssymb,
 reprint,%
]{revtex4-1}

\usepackage{graphicx}
\usepackage{dcolumn}
\usepackage{bm}

\usepackage[utf8]{inputenc}
\usepackage[T1]{fontenc}
\usepackage{mathptmx}
\usepackage{etoolbox}
\usepackage{hyperref}
\usepackage[version=4]{mhchem}
\usepackage{xr}
\def\NoComments{}

\ifdefined\NoComments

\newcommand{\comment}[1]{}
\else

\fi

\makeatletter
\def\@email#1#2{%
 \endgroup
 \patchcmd{\titleblock@produce}
  {\frontmatter@RRAPformat}
  {\frontmatter@RRAPformat{\produce@RRAP{*#1\href{mailto:#2}{#2}}}\frontmatter@RRAPformat}
  {}{}
}%
\makeatother
\begin{document}
\hyphenpenalty=50000
\tolerance=1000
\emergencystretch=3em

\preprint{AIP/123-QED}

\title{A Hierarchical Harmonic-to-Anharmonic Framework for \emph{ab initio} Screening and Mechanistic Analysis of Negative Thermal Expansion in  \texorpdfstring{\ce{ZrV2O7}}{ZrV2O7}}

\author{J. Bustamante}
\altaffiliation{%
These authors contributed equally to this work.}
\affiliation{%
Bundesanstalt für Materialforschung und -prüfung (BAM), Materials Chemistry, Berlin, Germany}
\affiliation{%
Freie Universit\"at Berlin, Department of Biology, Chemistry, Pharmacy, Berlin, Germany}

\author{A. Miliūtė}
\altaffiliation{%
These authors contributed equally to this work.}
\affiliation{%
Bundesanstalt für Materialforschung und -prüfung (BAM), Materials Chemistry, Berlin, Germany} 
\affiliation{%
The Humboldt University of Berlin, Department of Chemistry, Berlin, Germany
}

\author{N. Jaykhedkar}
 \affiliation{%
 Bundesanstalt für Materialforschung und -prüfung (BAM), Materials Chemistry, Berlin, Germany}

\author{A. A. Naik}
\affiliation{%
Bundesanstalt für Materialforschung und -prüfung (BAM), Materials Chemistry, Berlin, Germany}
\affiliation{%
Friedrich Schiller University of Jena, Institute of Condensed Matter Theory and Optics, Jena, Germany}

\author{J. Grandel}
\affiliation{%
Bundesanstalt für Materialforschung und -prüfung (BAM), Materials Chemistry, Berlin, Germany}
\affiliation{%
Friedrich Schiller University of Jena, Institute of Condensed Matter Theory and Optics, Jena, Germany}

\author{F. Emmerling}
\affiliation{%
Bundesanstalt für Materialforschung und -prüfung (BAM), Materials Chemistry, Berlin, Germany} 
\affiliation{%
The Humboldt University of Berlin, Department of Chemistry, Berlin, Germany
}

\author{T. M. Stawski}
\affiliation{%
Bundesanstalt für Materialforschung und -prüfung (BAM), Materials Chemistry, Berlin, Germany}

\author{J. George$^\dagger$}
\affiliation{%
Bundesanstalt für Materialforschung und -prüfung (BAM), Materials Chemistry, Berlin, Germany}
\affiliation{%
Friedrich Schiller University of Jena, Institute of Condensed Matter Theory and Optics, Jena, Germany}

\date{\today}

\begin{abstract}

Negative thermal expansion (NTE) in framework materials arises from complex lattice dynamics that are challenging to capture using conventional theoretical approaches.
Here, we establish a hierarchical computational strategy for investigating NTE in ZrV$_2$O$_7$, integrating the quasiharmonic approximation (QHA), temperature-dependent effective potentials (TDEP), and machine-learning-accelerated \textit{ab initio} molecular dynamics (ML-MD). Quasi-harmonic lattice-dynamical calculations identify low-frequency phonon modes with negative Gr\"uneisen parameters as inexpensive screening descriptors, while TDEP incorporates finite-temperature anharmonicity and ML-MD further reveals temperature-dependent structural distortions that drive the lattice contraction. The observations are further corroborated with X-ray diffraction (XRD) measurements.
The concurrent increase in transverse displacement ($u_{T}$) provides microscopic evidence for a quasi-rigid-unit-mode (QRUM) mechanism underlying the NTE behaviour at elevated temperatures. This is consistent with previous reports on other framework materials, including ZrW$_2$O$_7$, where collective polyhedral rotations and transverse atomic motions have been linked to NTE. 
Overall, our results demonstrate how relatively inexpensive lattice-dynamical descriptors can identify potential NTE materials, while higher-level anharmonic simulations can then resolve the underlying structural mechanisms at the atomic scale.

\end{abstract}

\maketitle

%

\section{Introduction}\label{intro}

\begin{figure*}
\includegraphics[scale=0.54]{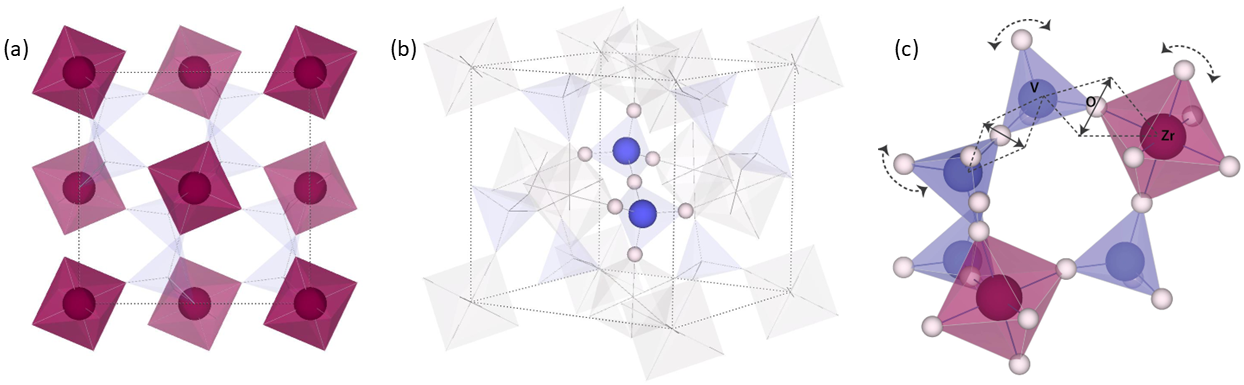}
\caption{(a) The ZrO$_6$ polyhedra in ZrV$_2$O$_7$ occupy the corners and face centers of the unit cell, arranged in an alternating pattern of clockwise and counterclockwise rotations along the \textit{c}-axis. The ZrO$_6$ polyhedra are not directly linked to one another; instead, they are connected through VO$_4$ tetrahedra through shared oxygen atoms, forming a corner-sharing framework. (b) At the center of the unit cell, two VO$_4$ tetrahedra are linked together, creating a V$_2$O$_7$ bridging unit that connects the surrounding ZrO$_6$ polyhedra. (c) Schematic representation illustrating how transverse displacement of the bridging oxygen triggers rotations of the VO$_4$ and ZrO$_6$ polyhedra, resulting in collective distortions in the framework.}
\label{fig:structure}
\end{figure*}

 Materials exhibiting negative thermal expansion (NTE) are of significant fundamental and technological interest due to their anomalous lattice response to temperature and their potential for thermal expansion compensation in composite systems.~\cite{li2026negative, evans1999} For instance, materials such as ZrW$_2$O$_8$~\cite{Hao2021} and ZrV$_2$O$_7$~\cite{Shreenivasa2019} are used to coat battery cathodes to reduce mechanical degradation and strain in batteries. While ZrW$_2$O$_8$ is among the most widely studied NTE materials,~\cite{Mary1996, Korthuis1995, EVANS1997, Evans1996, evans1999,Lind2012} ZrV$_2$O$_7$ has also attracted increasing attention.~\cite{Evans_1998, Withers1998, evans1999, Korthuis1995, Chen2024, Miliute2025} An interesting feature of ZrV$_2$O$_7$ is its isotropic NTE over a broad temperature range, from around 400 K all the way up to $\sim$1000 K.~\cite{Chen2024} It crystallizes in a cubic crystal structure with \textit{Pa$\overline{3}$} space group symmetry, consisting of an interconnected three-dimensional framework of corner-sharing ZrO$_6$ octahedra and VO$_4$ tetrahedra (see Fig.~\ref{fig:structure}). 

Earlier experimental reports demonstrate that it exists as 3$\times$3$\times$3 superstructure at low temperatures arising from a periodic arrangement of collective polyhedral tilts, corresponding to a lattice instability at the commensurate wavevector $q = (1/3, 1/3, 1/3)$.~\cite{Withers1998}
Experimentally, ZrV$_2$O$_7$ undergoes a structural transformation that manifests as a discontinuity (or anomaly) in the temperature–volume relationship, separating regions of distinct thermal expansion behavior. 
Withers et. al.~\cite{Withers1998,Korthuis1995} reported two successive phase transitions at 350 K and 375 K, based on differential scanning calorimetry (DSC) results, neutron and X-ray diffraction data and a detailed electron diffraction study.  However, the nature of this intermediate phase was characterized as a complex incommensurately modulated phase and remains not fully resolved.

A fair increase in the lattice parameters is observed at low temperatures ($\sim$300 K to 380 K), beyond which NTE behaviour sets in. The decrease in the lattice parameters is widely attributed to low-energy phonon modes involving correlated rotations and distortions of VO$_4$ tetrahedra and ZrO$_6$ octahedra.~\cite{KHOSROVANI1997} These quasi rigid-unit modes (QRUM), arising from the transverse vibrational motion of oxygen atoms, are considered to be the key mechanism driving the contraction of the crystal lattice upon further heating. In other words, heating activates low-frequency lattice modes, where flexible bridging linkages and collective transverse vibrations drive contraction, while accommodating changes in polyhedral shape.~\cite{KHOSROVANI1997, shi2021negative} Mechanisms involving QRUM and anharmonic effects were also observed in other framework materials. Studies on ScF$_3$ by Bucko \textit{et al.}~\cite{Bucko2015} show NTE arising from a competition between thermal expansion of Sc--F bonds and increasing octahedral tilting. ZrW$_2$O$_8$ analysis by Sanson \textit{et al.}~\cite{Sanson2014} indicates transverse vibrations of stiff Zr--O/W--O bonds and significant polyhedral deformation. Compared to these systems, which possess a relatively symmetric and coherent polyhedral framework that can be adequately described within primitive or small supercells, ZrV$_2$O$_7$ requires a larger supercell to accommodate its long-range structural modulation and cooperative QRUM motions.

Theoretical investigations of this system have historically been limited by a fundamental computational trade-off. Accurately describing the long-range collective polyhedral rotations requires large supercells, such as the experimentally observed superstructure containing 1080 atoms, yet the strong anharmonicity of the system demands explicit finite-temperature sampling that is prohibitively expensive at this scale using standard first-principles methods. Consequently, previous computational studies have largely restricted calculations to smaller primitive cells and relied on conventional harmonic approximations, inherently neglecting the dynamic structural fluctuations central to the NTE mechanism.~\cite{mittal_2008, Miliute2025} We therefore address two critical gaps in the existing literature:
(i) the lack of a fully anharmonic, atomic-level description of the NTE behavior of ZrV$_2$O$_7$ within an adequately sized supercell, and 
(ii) a detailed microscopic understanding, through computational modelling, of the structural changes driving these anomalies in its thermal expansion.
\begin{figure*}
\includegraphics[scale=0.50]{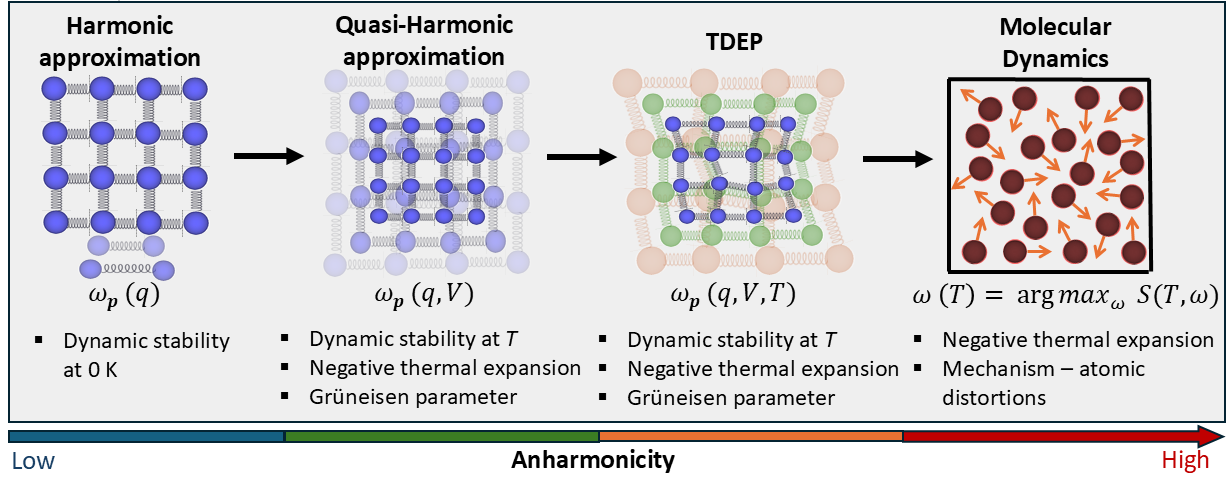}
\caption{Hierarchy of computational approaches for investigating lattice dynamics and anharmonicity. From left to right, the harmonic approximation accounts only for lattice vibrations at 0 K, the quasi-harmonic approximation incorporates volume-dependent phonons to capture thermal expansion, the temperature-dependent effective potential (TDEP) method includes finite-temperature anharmonic effects through renormalized phonons, and molecular dynamics (MD) explicitly samples anharmonic atomic motion, enabling the investigation of strongly anharmonic materials and dynamical processes. The degree of anharmonicity captured increases from left to right.}
\label{overview}
\end{figure*}

We, thus address the current gaps, by employing a hierarchy of computational methods, systematically recovering these missing anharmonic effects. To this end, in addition to the traditional QHA  and QHA$_{mod}$ approaches,~\cite{Stoffel2010, deringer_2014} which introduce volume dependency into harmonic phonon frequencies, the temperature-dependent effective potential (TDEP) method is employed. In this approach, effective interatomic force constants are extracted by fitting forces obtained from thermally sampled configurations, enabling the construction of a temperature-dependent Hamiltonian.~\cite{Bottin2020} Nevertheless, this approach can be insufficient in systems where higher-order anharmonicity, strong mode coupling, and dynamic structural disorder play a dominant role.  To address these limitations, we use \textit{ab initio} molecular dynamics (AIMD) that offers a natural framework for explicitly sampling anharmonic lattice dynamics and entropic contributions at finite temperature. Recent advances in machine-learned interatomic potentials (MLIPs) have further expanded the accessible length and time scales. Potentials trained on smaller unit-cell data can be employed to perform large-scale simulations with near \textit{ab initio} accuracy, rendering the investigation of such complex systems computationally feasible.

While the work follows a clear methodological progression, from harmonic to fully anharmonic descriptions (as shown in Fig.~\ref{overview}), highlighting the strengths and limitations of each approach in capturing the NTE behavior, it goes beyond a purely technical comparison. At each hierarchical level, from computationally inexpensive to increasingly sophisticated approaches, we identify robust structural descriptors that capture the essential features of NTE. These descriptors provide a physically motivated basis for screening large number of crystal structures and identifying promising NTE candidates at an appropriate computational cost. 

The paper is organized as follows: section II provides a detailed description of the experimental and computational methodology employed in this study. Section III presents the results and is divided into three parts ---QHA, TDEP, and AIMD--- through which the mechanism underlying NTE is progressively elucidated. Finally, the main findings and implications are summarized in the conclusions.

\section{Methodology}\label{methods}

\subsection{Structural, Electronic and Vibrational Analysis}\label{st_ele_vib_method}

Structure optimisation was performed using density functional theory (DFT) with the Perdew-Burke-Ernzerhof (PBE), PBE employed with Grimme's D3 dispersion correction,~\cite{Grimme2010,Grimme2011} and SCAN+rVV10 functionals~\cite{peng2016scan+} as implemented in the Vienna Ab initio Simulation Package ($VASP$).~\cite{kresse_ab_1993, kresse_theory_1994, Kresse1996,kresse_ultrasoft_1999} Our calculations used the projected augmented-wave method,~\cite{kresse_ultrasoft_1999,blochl_projector_1994} with a kinetic energy cut-off of 520 eV. For static relaxations, the conventional unit cell of the ZrV$_2$O$_7$ structure in the \textit{Pa$\overline{3}$} phase, containing 40 atoms each, has been used. All structures were relaxed until the maximum components of forces acting on atoms were smaller than 10$^{-3}$ eV/{\AA}. The total energy was strictly converged using a $\Gamma$-centre grid of (6$\times$6$\times$6) $k$-points and $\Delta$E $<$ 10$^{-8}$ and $\Delta$E $<$ 10$^{-6}$ eV for the electronic and structural optimisation, respectively. Based on the comparison of the structure optimization obtained with different exchange-correlation functionals and as a compromise with regard to computational cost, PBE+D3 was selected for the subsequent vibrational properties of the material. A supercell approach based on the finite displacement method, as implemented in \textsc{Phonopy},~\cite{togo_first-principles_2008,togo_first_2015,togo_implementation_2023} was used. A 2$\times$2$\times$2 (320 atoms) supercell with displacements of 0.01 {\AA} were tested to extract the harmonic force constants. The corresponding force calculations were performed using $\Gamma$-centered 3$\times$3$\times$3 k-point mesh. Finally, to correct for long-range dipole-dipole interactions, Born effective charge computations were also performed, and the non-analytical term correction, implemented within \textsc{Phonopy}, was applied.~\cite{gonze_dynamical_1997,gonze_interatomic_1994}

To gain further insight into the chemical bonding underlying the structural and vibrational characteristics of ZrV$_2$O$_7$, the Zr--O and V--O interactions were analyzed using the \textsc{LOBSTER} software through our automated bonding analysis workflow. Details of the sub-tasks in this workflow are available in the corresponding publication.~\cite{deringer2011crystal, maintz2013, maintz2016lobster, nelson2020lobster} The Zr--O and V--O interactions were analysed using the energy-integrated value up to the Fermi level of the Crystal Orbital Hamilton Population (ICOHP)~\cite{dronskowski1993} ---a quantitative descriptor of bond strength--- and energy-integrated Crystal Orbital Bond Index (ICOBI)~\cite{muller2021} ---a quantitative descriptor of bond order. The fully optimized structure from the phonon computations is used as the input structure in bonding analysis calculations. Post-processing of the \textsc{LOBSTER} outputs was performed using \textit{LobsterPy}.~\cite{george2022automated,naik2024lobsterpy}

\subsection{Quasi-harmonic approximation (QHA)}\label{qha_method}

Within the QHA, a series of seven regularly spaced volumes ($V$), referenced to the optimised ground-state V$_0$, ranging from 0.98$^3$$\times$$V_0$ to 1.04$^3$$\times$V$_0$ in increments of roughly 0.03$\times$V$_0$ was considered at PBE+D3 level. At each volume, the lattice parameters and atomic positions were relaxed, while keeping the cell volume fixed, followed by vibrational calculations using \textsc{Phonopy}. The corresponding Helmholtz free energy A$_{\mathrm{QHA}}$ ($T$,$V$) values was determined using the equation:
\begin{equation}
A_{\mathrm{QHA}}(T,V) = E_0(V) + k_BT\sum_{\textbf{q}p} ln\left[1-exp \left({\frac{-\hbar \omega_p(\textbf{q},V)}{k_BT}}\right) \right]
\label{eq:helmholtz_energy_1}   
\end{equation}
where $E_{0}(V)$ is the electronic energy and $\omega_p(T,V)$ is the harmonic frequency of the wavevector $\textbf{q}$ for the phonon mode $p$.
For each temperature ($T$), the resulting A$_{QHA}$($T$,$V$) vs $V$ relation was then fitted using the Vinet equation of state with the equilibrium free energy (A$_{QHA}$($T$,$V$$_0$($T$))), equilibrium volume ($V$$_0$($T$)), bulk modulus ($B$$_0$($T$)), and its pressure derivative [B$'_0$(T)] as the fitting parameters. The lattice parameters corresponding to $V_0$(T) were subsequently obtained to determine the thermal expansion.
Dynamical instabilities were encountered at contracted volumes within the conventional QHA, manifested by the appearance of imaginary phonon frequencies as shown in Fig. S1 in the Supplementary Information (SI). To account for these dynamical instabilities, a modified quasi-harmonic approximation (QHA$_{mod}$) approach following Deringer et al.'s work was employed.~\cite{Stoffel2010, deringer_2014, george_2016} The same volume-dependent electronic energy calculations and structural relaxations were retained, while the vibrational contribution $A_{vib} (T,V)$ at dynamically unstable volumes ($V\leq V_0$), was extrapolated from the dynamically stable volumes ($V\geq V_0$) using a linear fit, exploiting the approximately linear dependence of the $A_{vib}$ on volume. We then combined the extrapolated vibrational contribution with the DFT ground-state energy ($E_{0} (V)$) to obtain the modified Helmholtz free energy. The resulting free-energy curve was then used to determine the equilibrium volume ($V_{0} (T)$) at each temperature (Fig. S3) and thereby obtain the thermal expansion behaviour. For the expanded volume calculations, a (2$\times$2$\times$2) supercell was found to be sufficient to converge the vibrational contributions and associated thermal properties.

Additionally, mode-based Gr\"uneisen parameter, which is a measure of change in vibrational mode frequency with change in volume, given by the relation $\gamma_i = -\frac{V}{\omega_i} \frac{\partial \omega_i}{\partial V}$ were obtained. For this, two additional structural optimisations were performed at constant volume: $\pm$1\%V$_0$. However, the phonon dispersion calculated at $-1$\%V$_0$ exhibited imaginary modes. Therefore, a smaller contraction corresponding to $-0.5$\%V$_0$ was used to obtain a dynamically stable phonon spectrum and enable a reliable evaluation of the mode-resolved Gr\"uneisen parameters. Furthermore, average Gr\"uneisen parameters were also used to evaluate the volumetric thermal expansion coefficient within the quasiharmonic approximation, as described in Section S1.1 of the SI. 

\subsection{The Temperature Dependent Effective Potential (TDEP)}\label{tdep_method}

Following the QHA calculations, the stochastic Temperature Dependent Effective Potential (s-TDEP) method~\cite{Kim1997, Hellman2011, hellman2013temperature,hellman_2013,knoop2024tdep} was employed at the PBE+D3 level to account for $T$-induced anharmonic effects on the phonon spectra. In this approach, atomic configurations representative of the canonical ensemble ($NVT$), in which the number of atoms ($N$), volume ($V$), and temperature ($T$) are fixed, were generated via stochastic sampling over a temperature range of 300 -- 800 K (in steps of 100 K) and volumes in the range 0.99$^3$$\times$ $V_0$ $\leq$ V $\leq$ 1.02$^3$ $\times$ $V_0$ with increments of roughly 0.03 $\times$ $V_0$, using a 3$\times$3$\times$3 supercell (1080 atoms). The stochastic sampling is performed within an iterative self-consistent loop, ensuring configurations representative of thermodynamic equilibrium at each target $T$.~\cite{ladygin2025anharmonic} Convergence of both the free energy and the renormalised phonon frequencies was achieved after nine iterations, corresponding to a total of 1024 sampled configurations. The free energy difference between the last two iterations was below 10$^{-4}$ eV, indicating that the self-consistent cycle had reached convergence.  The cutoff radius was set to 10 {\AA} for obtaining the second-order force constants, as further increase in the cutoff or number of iterations had negligible impact on free energy convergence.
Due to the large number of sample configurations, direct DFT evaluation of forces was computationally prohibitive. Therefore, forces were computed using machine-learning interatomic potentials (MLIP) trained at 300 K using on-the-fly learning as implemented in \textsc{VASP} (described in the next Section II C). 

Within this framework, the Helmholtz free energy was evaluated using the renormalized $T$-dependent force constants across the considered volume range as given below;
\begin{equation}
A_{\mathrm{TDEP}}(T,V) = E_0(V) + k_BT\sum_{\textbf{q}p} ln\left[1-exp \left({\frac{-\hbar \omega_p(\textbf{q},V,T)}{k_BT}}\right) \right]
\label{eq:helmholtz_energy_2}   
\end{equation}
For each temperature, the Helmholtz free energy was evaluated over the selected range of volumes using the temperature-renormalized force constants. The equilibrium volume was then identified as the volume that minimizes the free energy. The resulting free energy–volume curves were fitted using the Vinet equation of state, consistent with the procedure adopted in QHA as shown in section S1.1 of the SI.
The formal expression for the vibrational free energy remains identical in both the QHA and the TDEP framework; the key difference lies in how the phonon frequencies are obtained. Within QHA, phonons are treated harmonically at each fixed volume, and anharmonic effects are therefore incorporated only implicitly through the volume dependence of the phonon frequencies, such that $(\omega_p=\omega_p(V(T)))$. Consequently, their temperature dependence arises through the temperature-dependent equilibrium volume. In TDEP, on the other hand, they are derived from temperature-renormalized interatomic force constants, leading to an explicit dependence on both temperature and volume, $\omega_p = \omega_p (T,V(T))$. This temperature dependence arises from the underlying anharmonic potential energy surface sampled at finite temperatures.

\subsection{Ab-initio molecular dynamics computations (AIMD) }\label{aimd_method} 

Ab initio molecular dynamics (AIMD) in the $NPT$ ensemble at zero pressure and temperatures of 300 K, 400 K, 500 K, 600 K, 700 K, and 800 K were performed for 3$\times$3$\times$3 supercells (1080 atoms) using $VASP$.~\cite{kresse_ab_1993} The atomic positions and lattice parameters were treated as dynamic variables. The simulation temperature was maintained using a Langevin thermostat~\cite{Lemons1997, Tuckerman_5} with friction coefficients of 2 ps$^{-1}$ and 10 ps$^{-1}$ for atomic and lattice degrees of freedom, respectively. The pressure was controlled by a Parrinello-Rahman barostat,~\cite{Parrinello1980, Parrinello1981} with the mass associated with the lattice degrees of freedom set to 2 amu. The velocity Verlet algorithm~\cite{Frenkel_Smith} with a time step of 1 fs was used to integrate the equations of motion of the ions. Consistent with the setting used in the relaxations, a single $k$-point grid was used for the large supercells in all MD simulations. To account for potential long-range dispersion interactions, corrections were employed using the zero-damping DFT+D3 method proposed by Grimme \textit{et al.}~\cite{Grimme2010, Grimme2011} 

To accelerate MD simulations, on-the-fly machine learning was employed, as recently implemented in $VASP$.~\cite{Jinnouchi2019, Jinnouchi2020, Kermode:2015} For the description of the machine-learned potential energy surface, descriptors based on the Gaussian representation of the atomic distribution,~\cite{Jinnouchi2020,Behler2007} were used. The predicted target properties, i.e., energies, forces, and stress tensor components, were obtained via Bayesian linear regression, allowing for reliable estimates of prediction quality.~\cite{Jinnouchi2019,Jinnouchi2020, Behler2007} A separate training procedure was executed for each temperature considered. The training was performed on a 2$\times$2$\times$2 supercell (320 atoms) for at least 60 ps at each temperature, ensuring sufficient structures ($\ge 50$) were collected in the training data set. Re-fitting the training data using several hyperparameters showed that the regularization term ($\sigma_w$) had the strongest impact on reducing errors, and was therefore fixed at 10$^{-4}$. After refitting, the MLIP model reached training errors of 10$^{-4}$ eV, 10$^{-2}$ eV/{\AA} and 10$^{-1}$ eV/{\AA}$^2$ for energies, forces and stresses respectively. Post-training, a production run was performed using the MLIP for the 3$\times$3$\times$3 supercell, generating a trajectory of 1 ns. To validate the accuracy of the trained force fields for the predicted MD trajectories, 40 random structures were sampled and evaluated with single-point DFT to obtain reference energies and forces. These values were compared against the MLIP predictions, which closely follow the reference DFT values with minimal scatter, as shown in the parity plots in Fig. S7 of the SI. The resulting RMSE values for energies and forces across all temperatures remain within 0.001 eV/atom and 0.01 eV/{\AA}, respectively, as reported in Section S3 of the SI. Further validation was carried out by comparing the phonon band structures predicted by the MLIP with those obtained from DFT, as shown in Fig. S5 of the SI.
After obtaining the production data from the MLIP-MD simulations, an appropriate equilibration length was determined using statistical Mann-Kendall tests,~\cite{Schiferl1985} and the corresponding data were discarded. The basic observables, such as the lattice parameters, volume, radial distribution function (RDF), and distribution of bond angles, were computed as averages over the equilibrated MD trajectory. The corresponding standard error was calculated using the blocking method.~\cite{Flyvbjerg1989}

The internal deformation of the ZrO$_6$ and VO$_4$ polyhedra or henceforth referred to as intra-polyhedral deformation was quantified using the continuous symmetry measure (CSM), as implemented in \textsc{ChemEnv}.~\cite{Waroquiers2020, ong2013python, george_pymatgen_2026} CSM measures the deviation of an observed polyhedron from its corresponding ideal reference geometry, which is defined in \textsc{ChemEnv} as a perfect polyhedron. A CSM of 0 corresponds to an ideal, undistorted polyhedron, while increasing values (0–100) indicate increasing deviation from the reference shape. The measure is obtained by optimally matching the observed and ideal polyhedra and quantifying their normalized structural deviation. 

In addition, the transversal displacement ($u_T$), as described in Jaykhedkar \textit{et al.,}~\cite{Jaykhedkar:2023} was calculated as an effective metric to quantify inter-polyhedral distortions, i.e the deviation of the interconnections between ZrO$_6$--VO$_4$ and VO$_4$--VO$_4$, polyhedra from their ideal geometries. This quantity measures the extent to which an atom is displaced perpendicular to the ideal bond direction within the polyhedral framework. Thus, the transversal displacement $u_T$, in this study, is defined as the perpendicular distance between the displaced oxygen atom and the straight line joining, for instance, the Zr and V atoms of the ZrO$_6$ octahedra and VO$_4$ tetrahedra, respectively (see Figure~\ref {fig:structure} (c) for clarity). Using the O--V--Zr angle ($\theta$) and the O--V bond length ($r_{O-V}$), $u_T$ can be calculated as $u_T = r_{O-V}.sin \theta$. Thus, CSM and $u_T$ provide quantitative measures of intra-polyhedral deformation and inter-polyhedral distortion, respectively, in the ZrO$_6$--VO$_4$ framework.

\subsection{Synchrotron experiments}\label{synchrotron_method} 

A single-phase ZrV$_2$O$_7$ sample was synthesized via hydrolytic sol-gel synthesis described elsewhere.~\cite{Miliute2025} High-temperature high-energy X-ray diffraction (HEXD) measurements of ZrV$_2$O$_7$ were performed at the DESY PETRA III P02.1 Powder Diffraction and Total Scattering Beamline over the 25–810 K temperature range and a q range of 0.0015 to 5.47 {\AA}$^{-1}$. X-rays with a fixed energy of 60 keV and a wavelength of approximately $\lambda \sim$ 0.207 {\AA} were used. The diffraction patterns were collected using a Varex XRD 4343CT area detector (2880 $\times$2880 pixels; 150 $\times$150 $\mu$m$^{2}$ pixel size) with a sample-to-detector distance (SDD) of 2000 mm. The sample was contained in a quartz capillary with an internal diameter of 1.0 mm that was continuously rotated during acquisition to improve powder averaging. The sample was heated with a hot-air blower. Two-dimensional diffraction images were processed using DAWN Science, with detector geometry calibrated using a LaB$_6$ standard. The calibration determined the beam centre, sample-to-detector distance, and detector tilt. Masks were applied to exclude defective detector pixels and the beamstop region, and empty-capillary measurements collected under identical conditions were used for background correction. The corrected 2D images were azimuthally integrated to obtain one-dimensional diffraction patterns as a function of the scattering angle or the momentum transfer, Q, for subsequent structural analysis (see Fig. S9) of the SI). The lattice parameter was extracted by indexing the first 15 peaks of 3.1 \% minimum intensity via the Treor indexing method.

\section{Results and Discussion}\label{results}

\subsection{Structural, Electronic and Vibrational Analysis}\label{st_ele_vib_res}

We begin by statically relaxing the 40-atom unit cell of ZrV$_2$O$_7$. To assess the influence of the exchange-correlation functionals, we evaluated PBE, PBE$+$D3, and SCAN$+$rVV10. All three functionals predict lattice parameters within 3\% of the experimental value, as summarized in Table~\ref{tab:table1}. 
\begin{table}[!ht]
\begin{ruledtabular}
\centering
\begin{tabular}{lcr}
Functional & a ({\AA}) & V ({\AA}$^3$)\\
\hline
\\
Exp$_{(300~K)}$ & 8.779 & 676.8 \\
PBE & 8.99 (2.4\%) & 725.96  \\
PBE+D3 & 8.96 (2.1\%) & 718.3 \\
SCAN+rVV10 & 8.88 (1.1\%) & 701.17 \\
{Evans \textit{et. al.}~\cite{Evans_1998}}& 8.765 & 673.1 \\
{Yamamura \textit{et. al.}~\cite{Yamamura2011}}& 8.783 & 677.6 \\
\end{tabular}
\end{ruledtabular}
\caption{\label{tab:table1} Calculated lattice parameters and volumes of the unit cell using the PBE, PBE+D3, and SCAN+rVV10 functionals, compared with in-house experimental measurements and previously reported values. The values in parentheses represent the percentage overestimation relative to the in-house experimental results.}
\end{table}

\begin{figure*}[!ht]
\includegraphics[width=\linewidth]{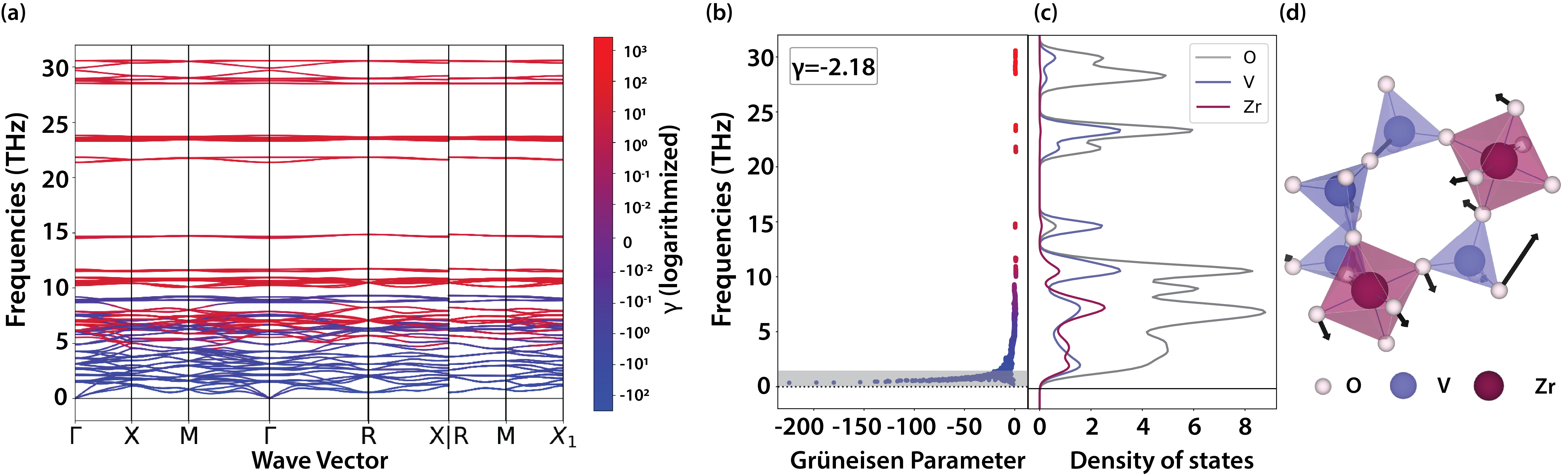}\\
\caption{Phonon properties of the parent ZrV$_2$O$_7$ structure: (a) Phonon dispersion curve along the high-symmetry path in the Brillouin zone, colored according to the mode-resolved Gr\"uneisen parameter, (b) mode-based Gr\"uneisen parameter plot, and (c) partial phonon density of states (PDOS). The absence of imaginary frequencies confirms the dynamical stability of the cubic phase at the equilibrium volume. The phonon spectrum exhibits distinct low- and high-frequency vibrational regions separated by a pronounced phonon gap between approximately 15 and 23 THz, characteristic of the vibrational spectrum commonly observed in framework NTE materials. The PDOS shows that O and V atoms contribute broadly across the vibrational spectrum, while Zr vibrations are mainly concentrated in the low-frequency region. Mode-based Gr\"uneisen parameters show that the low-energy vibrational modes (highlighted in grey) exhibit strong anharmonicity, as evidenced by large Gr\"uneisen parameters. The non-squared average Gr\"uneisen parameter is $\overline{\gamma}=-2.18$, highlighting the predominance of the negative Gr\"uneisen parameters associated with NTE. Colors in the Gr\"uneisen plot indicate band indices with ascending order of phonon frequencies. (d) Phonon eigenvectors associated with imaginary modes emerging upon volume contraction at $R$ point of the Brillouin zone ($\mathbf{q}=(0.5,0.5,0.5)$, in reciprocal lattice coordinates). These modes correspond predominantly to the collective O vibrations on the octahedra O--Zr--O and tetrahedral O--V--O. }
\label{fig:QHA}
\end{figure*}

Based on this agreement, we selected PBE$+$D3 as the primary level of theory for the subsequent QHA, QHA$_{mod}$, TDEP, and AIMD calculations, as it provides a reasonable compromise between accuracy and computational cost. Using a consistent level of theory across all methods ensured direct comparison between the different approaches. To assess the sensitivity of the results to the exchange-correlation functional, alternative functionals for structural optimization were also tested. However, due to its high computational cost, we could perform AIMD simulations at only one level of theory; therefore, the possible influence of the functional on the MD results cannot be ruled out.

\begin{figure}[!h]
\includegraphics[scale=0.5]{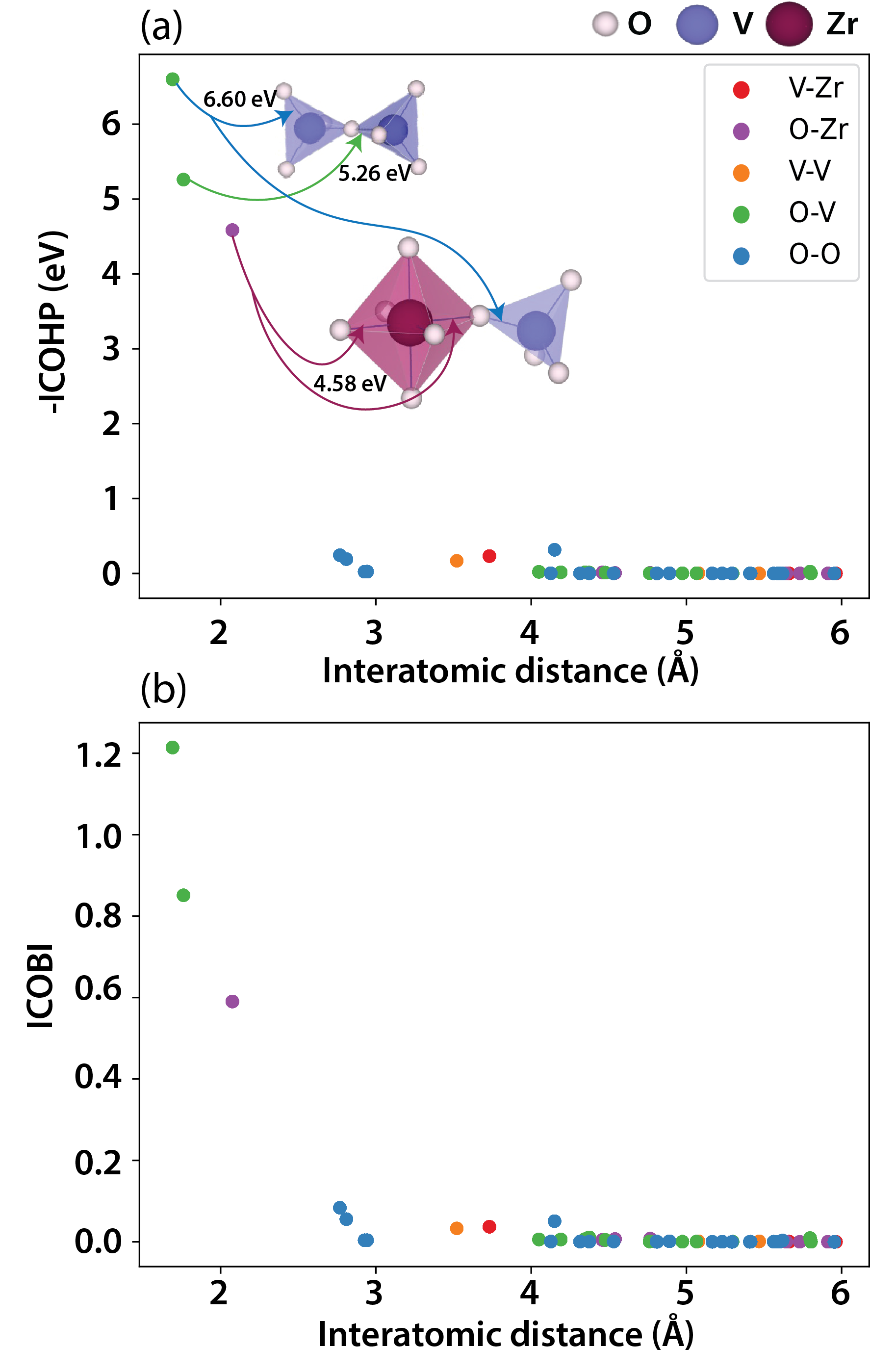} \\
\caption{ICOHP and ICOBI vs interatomic distances for ZrV$_2$O$_7$. V--O bonds within the VO$_4$ tetrahedral units and V--O interactions involving the bridging O atoms connecting neighbouring tetrahedra are shown. The Zr--O interactions correspond to bonds within the ZrO$_6$ octahedral units. The corresponding integrated values (ICOHP and ICOBI) quantify the bond strength and bond order, respectively, revealing strong interactions with pronounced covalent character throughout the polyhedral framework.}
\label{fig:ICOHP}
\end{figure}

As our first assessment of dynamical stability, vibrational analysis was performed using the harmonic approximation. The harmonic phonon frequencies along the high‐symmetry directions of the Brillouin zone show no imaginary modes, indicating the dynamical stability of the relaxed geometry at 0 K (Fig.~\ref{fig:QHA} (a)). This dynamically stable and optimized structure was then used for bonding analysis as presented in Fig.~\ref{fig:ICOHP}. Two distinct V--O bond types can be identified: the V--O bonds within the VO$_4$ tetrahedral units and the V--O interactions involving the bridging O atoms that connect neighboring tetrahedra. Both interactions are characterized by strong bonding, with ICOHP values of approximately $-$6.6 and $-$5.26~eV, respectively. Similarly, the Zr--O interactions display a large negative ICOHP $-$4.58~eV confirming strong bonding within the ZrO$_6$ polyhedral framework. The corresponding ICOBI values further support this. Approximately 1.2 and 0.8 are obtained for the V--O bonds within the tetrahedra and the bridging V--O interactions, respectively, whereas the Zr--O bonds exhibit ICOBI values of approximately 0.6, revealing pronounced covalent character in both types of interactions. Typical covalent single bonds have an ICOBI (bond order) of around 1. These results indicate that the V--O bonds are more covalent than the Zr--O bonds. Strong intra-polyhedral Zr--O and V--O covalent interactions are expected to support the robustness of the ZrO$_6$ and VO$_4$ units, allowing the framework to accommodate collective rotational and vibrational motions with minimal deformation of the individual polyhedra, as the CSM analysis presented in Section D will further corroborate.

Materials exhibiting NTE behaviour are commonly characterized by three features: (i) high-frequency phonon modes, (ii) a pronounced phonon gap that separates the high-energy phonon modes, and (iii) low-energy phonon modes with negative Gr\"uneisen parameters;~\cite{Yamamura2011,takenaka2012,Lind2012} which are evidently observed in Figure~\ref{fig:QHA} (a) and (b). A pronounced phonon gap is observed between approximately 15 and 23 THz in Fig.~\ref{fig:QHA} (a), separating low-frequency vibrations from high-frequency modes. The low-energy modes are particularly relevant to NTE because they can involve transverse and collective motions of the polyhedral framework that promote lattice contraction upon heating.~\cite{Yamamura2011,takenaka2012,Lind2012}  The corresponding partial phonon density of states (PDOS) (Fig.~\ref{fig:QHA} (c)) provides further insight into the atomic contributions to the vibrational spectrum. O and V contribute across a broad frequency range, with O vibrations dominating the low-frequency modes, whereas Zr contributes predominantly at lower frequencies (0--10 THz), consistent with its larger atomic mass and weaker Zr--O bond. Overall, these phonon-spectrum features are consistent with vibrational characteristics commonly reported for NTE materials and motivate further analysis of mode-resolved Gr\"uneisen parameters and their contributions to the NTE response. Moreover, these characteristic phonon-spectrum features may provide useful vibrational descriptors for identifying NTE materials, making them particularly relevant for future high-throughput screening strategies.

\subsection{Quasi-harmonic approximation}\label{qha_res} 

The harmonic analysis discussed above revealed the characteristic vibrational features commonly associated with NTE materials, particularly the presence of low-energy phonon modes separated from the high-frequency vibrations by a pronounced phonon gap. To assess whether these low-energy modes are directly involved in the NTE response of ZrV$_2$O$_7$, their volume dependence was further investigated through the mode-resolved Gr\"uneisen parameters within the QHA framework. For this purpose, structures were generated by systematically contracting and expanding the equilibrium volume $V_0$ in $\pm 1\%$ increments. However, even a slight compression of $-1\% V_0$ led to the emergence of negative phonon frequencies, indicating the onset of dynamical instability under compressed conditions (see Fig.S1 (b) and (c)). Even though moderate compressions introduced imaginary phonon frequencies, all expanded volumes remained dynamically stable with no imaginary modes observed. To ensure the correct prediction and avoid numerical artifacts, a slightly smaller contraction of 0.9985$^3$$\times$$V_0$(corresponding to $\sim-0.5\% V_0$) was adopted. In Fig.~\ref{fig:QHA} (b), we present the mode-based Gr\"uneisen parameter as a function of frequency.

The sign of the Gr\"uneisen parameter determines how vibrational frequencies respond to changes in volume. Positive values indicate that a mode stiffens upon compression and softens upon expansion, which is the conventional behaviour associated with positive thermal expansion. In contrast, negative values indicate anomalous behavior, with modes softening upon compression and stiffening upon expansion. 
The negative Gr\"uneisen parameters of the low-frequency modes in ZrV$_2$O$_7$ (highlighted in grey in Fig.~\ref{fig:QHA} (b)) therefore provide a direct physically meaningful indicator of the NTE behaviour. Moreover, their large negative values indicate a strong volume sensitivity of these low-energy modes, reflecting pronounced anharmonicity and structural flexibility.
Analysis of the corresponding phonon eigenvectors provides a structural interpretation of this anharmonic behavior. The low-energy modes with the most pronounced negative Gr\"uneisen parameters are primarily associated with librational motions~\cite{miller2009negative, ramirez1998, Yamamura2011} involving the Zr--O--V linkages, corresponding to collective rotations of the interconnected ZrO$_6$ octahedra and VO$_4$ tetrahedra, primarily accommodated by transverse displacements of the bridging O atoms. The internal geometries of the individual ZrO$_6$ and VO$_4$ units remain almost undistorted. These motions are consistent with the quasi-rigid unit mode (QRUM) mechanism previously discussed by Miller.~\cite{miller2009negative} 

Thus, a largely negative Gr\"uneisen value, particularly for low-frequency phonon modes, can serve as an initial indicator for screening materials that may exhibit NTE behaviour. Since these modes often dominate the vibrational entropy, their softening under compression suggests the lattice favors contraction upon heating. However, this criterion alone is not sufficient. Reliable identification of NTE behavior requires evaluating the full phonon spectrum, mode contributions, and anharmonic effects rather than relying solely on the presence of negative Gr\"uneisen values in low-frequency regions.~\cite{li2026negative}

To this end, the full phonon spectrum was systematically evaluated for both contracted and expanded equilibrium volumes in increments of roughly $\pm 3\%$, spanning a total volume variation of $\sim \pm 12\%$. As noted earlier, imaginary frequencies were observed for volumes $-1\%$ onwards. Further inspection of the eigenvectors (Fig.~\ref{fig:QHA} (d)) showed that these unstable modes primarily correspond to transverse oxygen motions connecting the ZrO$_6$ octahedra and VO$_4$ tetrahedra. Consequently, the standard QHA workflow could not be reliably applied to compute the thermal expansion coefficient, since the compressed volume states would need to be excluded from the analysis. 

To address this limitation, a modified quasi-harmonic approximation approach (QHA$_{mod}$) was adopted, as outlined in Section II.A, to determine the equilibrium volume at each temperature, thereby enabling the evaluation of thermal expansion. Although this procedure is an approximation, it effectively reconstructs the underlying free energy landscape and remains physically justified in the present case. The imaginary phonon mode arises from limitations of the harmonic approximation rather than indicating a true thermodynamic instability at finite temperature. Furthermore, in the same framework, as a complementary approach, the volumetric thermal expansion coefficient was evaluated using the temperature-dependent average Gr\"uneisen parameter (Refer to section S1.1  of the SI for further details). Together, these methods offer a computationally efficient first estimate of the thermal expansion behavior.

\begin{figure}[!h]
\includegraphics[scale=0.4]{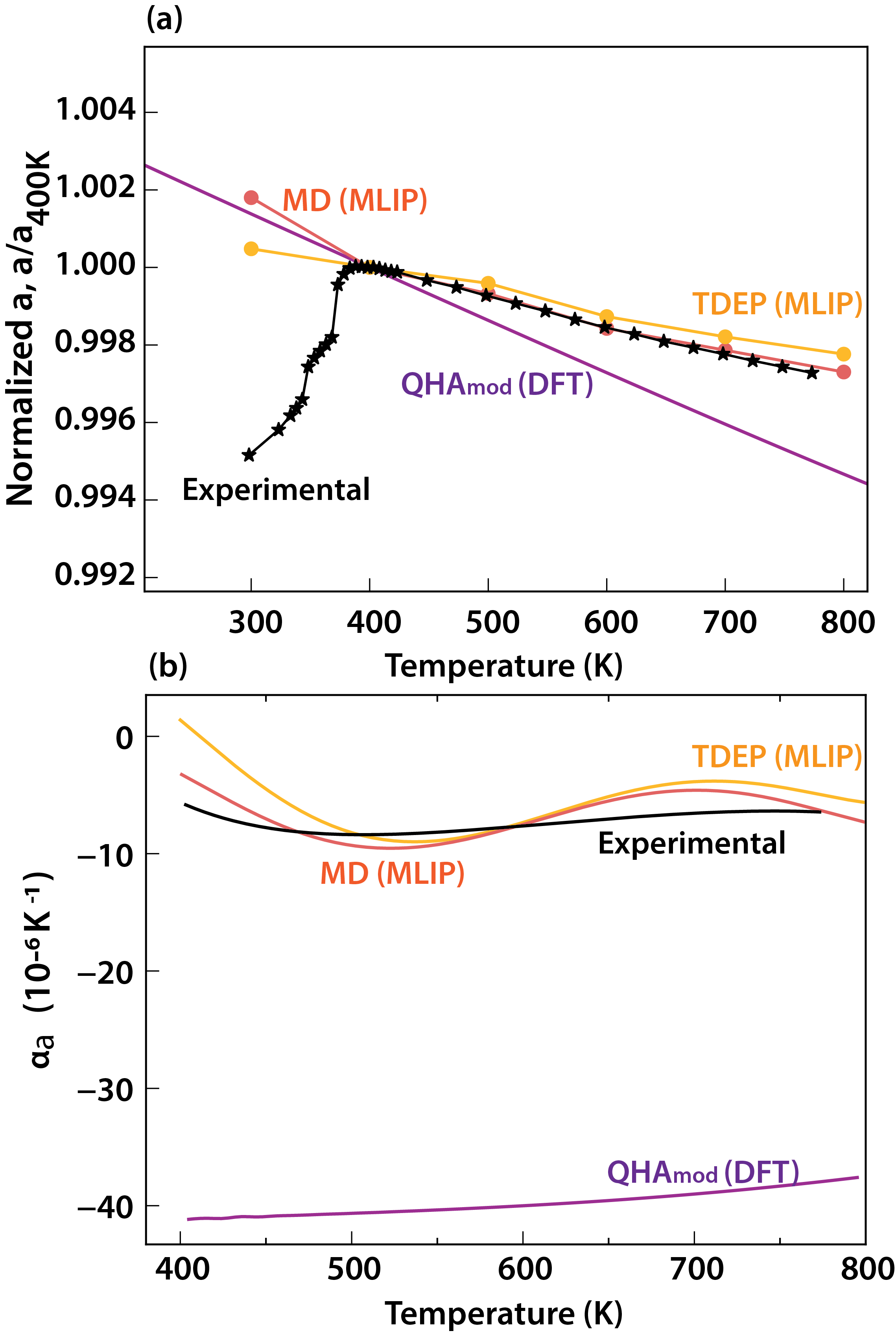} \\
\caption{ Temperature dependence of the normalized lattice parameter $a/a_{400}$ obtained from experiment, QHA, TDEP, and AIMD simulations. Note that $a_0$ corresponds to the value from experiments at 400 K. (b) The corresponding linear thermal expansion coefficient ($\alpha_a$) as a function of temperature predicted by the different methods is compared with experimental data.}
\label{fig:NTE}
\end{figure}

The temperature dependence of the normalized lattice parameter obtained using the QHA$_{mod}$ exhibits NTE over the investigated temperature range, but with markedly different magnitudes and temperature dependence as compared to experiment. In Fig.~\ref{fig:NTE} (a), the QHA prediction shows a nearly linear decrease in the normalized lattice parameter $a/a_{400}$ with increasing temperature, indicating persistent NTE. However, the corresponding lattice thermal expansion coefficient as shown in Fig.~\ref{fig:NTE} (b) is strongly negative ($-$40 ppm/K), which is substantially larger in magnitude than the experimental values. Experimentally, the lattice parameter begins to decrease from 400 K; however, above approximately 500–550 K, the contraction progressively weakens with increasing temperature. This is reflected in the experimental thermal expansion coefficient, which evolves from approximately $-$10 ppm/K at lower temperatures toward less negative values at higher temperatures.

Thus, while QHA correctly captures the negative sign of the thermal expansion coefficient (Fig.~\ref{fig:NTE} (b) and Fig. S2), hence the qualitative NTE behaviour, it significantly overestimates the magnitude of thermal contraction and does not reproduce its temperature dependence, highlighting the importance of anharmonic effects beyond the QHA. 
 Nevertheless, owing to its computational efficiency, this modified approach is a practical and inexpensive way for rapidly estimating thermal expansion behaviour in systems where conventional QHA fails due to phonon instabilities. However, it should be noted that it does not explicitly capture higher-order anharmonic effects, temperature-induced structural rearrangements, or dynamic processes such as polyhedral rotations that may contribute to NTE.

\subsection{TDEP with machine learning force field}\label{tdep_res}

To overcome some of the limitations of the QHA$_{mod}$ approach, TDEP was subsequently employed to incorporate explicit temperature dependence into the phonon spectra by fitting effective force constants to thermally sampled configurations. Although the Hamiltonian is truncated at the harmonic level, similar to QHA, the fitted second-order force constants are effectively renormalized by higher-order anharmonic interactions. In other words, the force constants are obtained from configurations with atomic displacements representative of finite temperatures. As a result, they inherently incorporate the effects of higher-order terms in the Taylor expansion of the potential energy, thereby accounting for anharmonicity. 

Phonon dispersion curves calculated at the ground state from TDEP at 300 K and QHA at 0 K are compared in  Fig. S4. In this case, the TDEP calculations give slightly softer phonon frequencies in the higher optical modes than the QHA results. This softening originates from the finite-temperature anharmonic potential energy surface captured within the TDEP framework and is therefore not an artifact of the MLIP. At contracted volumes, however, some differences emerge: while QHA  Fig.S1(b) predicts the appearance of imaginary phonon modes, TDEP stabilises these vibrational modes, enabling access to dynamically stable phonon spectra even under compression, as can be observed in Fig.S4(b). 

By comparing the free-energy profiles obtained from the  QHA$_{mod}$ and TDEP approaches in Fig.S3 and S6 in the SI, respectively, we observe a similar overall trend, with minor deviation in the equilibrium volume relative to the Vinet equation-of-state fit.

Above 400 K, TDEP accurately reproduces the experimentally observed decrease in the lattice parameter with increasing temperature, thereby capturing the NTE behavior of the material. Compared with QHA$_{mod}$, which substantially overestimates the magnitude of the thermal contraction, TDEP provides a significant improvement by yielding a much closer temperature dependence of the normalized lattice parameter, $a/a_{400}$, as evidenced by the more comparable slope to the experimental data. However, it systematically overestimates the normalized lattice parameter, as can be seen from the difference in the slope after 400 K. 

Moreover, $\alpha_a$ varies non-linearly and changes curvature as different phonon modes become thermally populated. At low temperatures, up to $\sim$500 K, the contraction becomes progressively stronger, with the minimum corresponding to $\sim$500-550 K where the contraction rate is largest. Beyond 600 K, the contraction becomes progressively weaker i.e. the lattice still contracts but at a slower rate. This could be attributed to the high frequency phonon modes, that favor expansion, becoming dominant and reducing net contraction. A pronounced deviation is observed in the curvature of the TDEP-derived $\alpha_a$ compared to experiment, arising from limitations in capturing higher-order anharmonic effects.
While the incorporation of anharmonicity through temperature-renormalized force constants in TDEP improves the phonon spectrum by eliminating imaginary modes and the NTE curve, the framework still relies on an effective harmonic Hamiltonian. As a result, it remains limited in capturing the strongly anharmonic and collective distortions that govern NTE. 

This indicates that, despite its improved treatment of lattice dynamics, TDEP may not be sufficient to fully describe the complex distortion mechanisms underlying thermal expansion in ZrV$_2$O$_7$.
Nevertheless, the qualitative agreement of NTE curves between experiment and TDEP supports its utility as an efficient screening tool for identifying potential NTE materials.

\subsection{ML-AIMD}\label{aimd_res}
Owing to the inherent limitations of the QHA and TDEP approaches in describing strongly anharmonic lattice dynamics, MD simulations accelerated by MLIPs, were employed to provide a more robust treatment of the thermal expansion behaviour in ZrV$_2$O$_7$, as described in Section II.C. 

The normalized lattice parameter ($a/a_{400}$) obtained from AIMD, shown in Fig.~\ref{fig:NTE} (a), exhibits further improvement over TDEP and is in better agreement with experiments, as evident from the nearly overlapping NTE curves beyond 400 K. Thus, the progressive incorporation of anharmonicity, from QHA through TDEP to ML-MD, is reflected in the increasingly close agreement of the predicted NTE behaviour with experiment. 
Despite good agreement at high temperatures, it fails to reproduce the initial lattice expansion at lower temperatures, as with TDEP and QHA. This limitation may reflect a different mechanism or a phase not captured at the current level of theory. 
\begin{figure}[!h]
\includegraphics[scale=0.29]{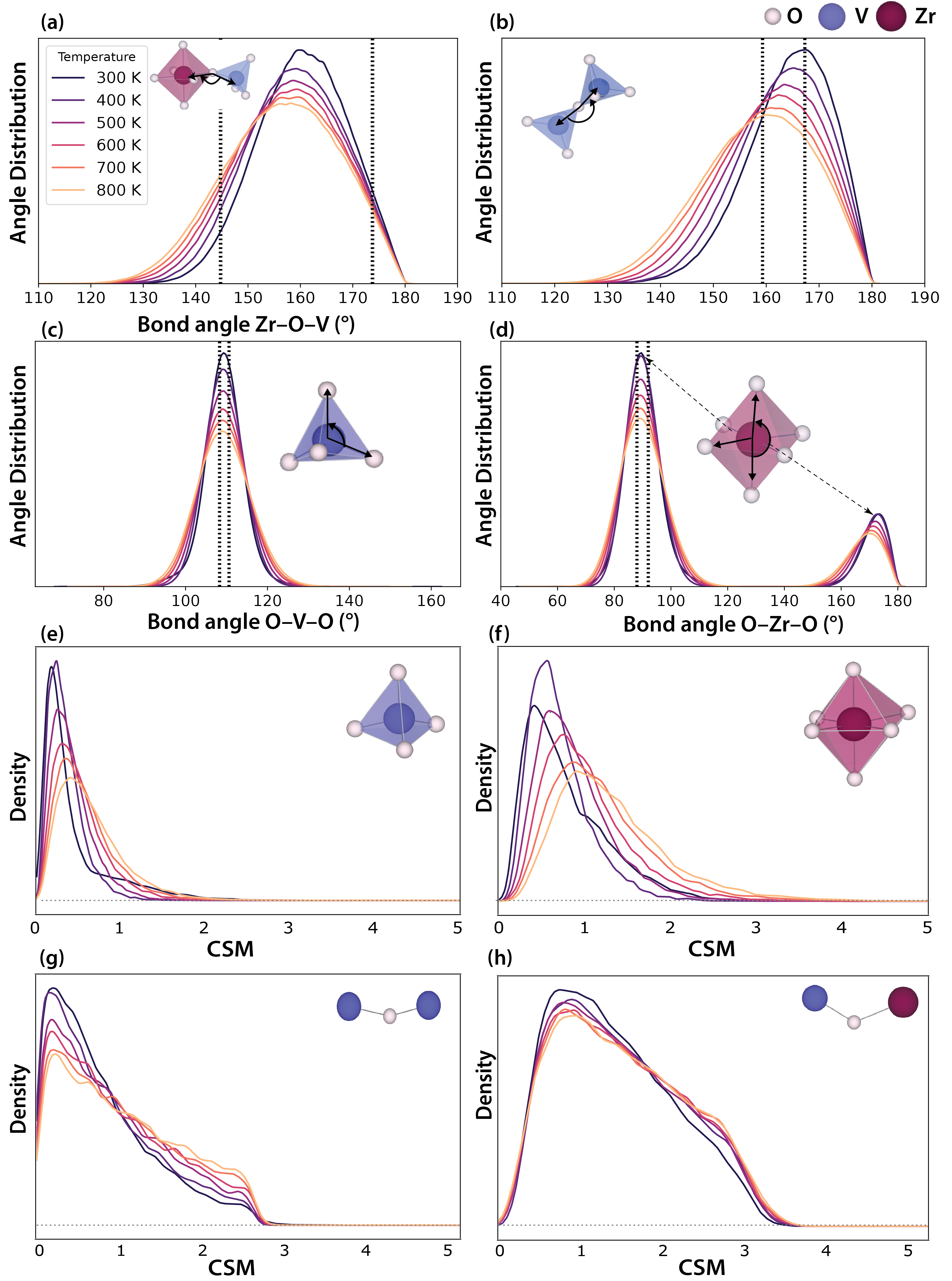}\\
\caption{Temperature-dependent distributions of bond angles for inter-polyhedral linkages, (a) Zr--O--V and (b) V--O--V, and intra-polyhedral angles within the coordination polyhedra, (c) O--V--O and (d) O--Zr--O. The geometric definitions of the corresponding angles are shown schematically in each panel. Distribution of the continuous symmetry measure (CSM) value for the (e) VO$_4$, (f) ZrO$_6$ polyhedra, (g) V--O--V and (h) Zr--O--V connections at different $T$. The gray dotted lines indicate the experimental bond-angle ranges reported by Evans \textit{et al.}~\cite{Evans_1998} at 300K: for the intra-polyhedral angles 108.3--110.6$^\circ$ for O--V--O, 88.0--92.0$^\circ$ and O--Zr--O, as well as inter-polyhedral linkage angles 159.3--167.6$^\circ$ for V--O--V, and 145--174$^\circ$ for Zr--O--V. }
\label{fig:angles}
\end{figure}
We infer an inherent underbinding in the theoretical description, which manifests as an overestimation of the equilibrium cell volume at 0 K. As summarized in Table~\ref{tab:table1}, including dispersion corrections in PBE only partially mitigates its well-known tendency to overestimate lattice parameters. In contrast, a more accurate meta-GGA functional such as SCAN, combined with nonlocal van der Waals corrections (rVV10), provides a better description of bonding and significantly reduces the deviation from experiment.
The impact of this under-binding is directly reflected in the internal atomic distortions. The larger theoretical lattice volume allows enhanced transverse displacements and framework distortions. Consequently, the simulations predict a stronger contraction than observed experimentally up to 400 K. In contrast, the smaller experimental lattice volume restricts these displacements, favouring thermal expansion over contraction.

A similar trend is evident in the corresponding thermal expansion coefficients ($\alpha_a$). As shown in Fig.~\ref{fig:NTE} (b), the AIMD-derived thermal expansion coefficient shows the closest agreement with the experimental data, outperforming both QHA and TDEP. In particular, it reproduces the temperature corresponding to the maximum contraction (530-540 K) more accurately and shows a less pronounced temperature dependence, resulting in reduced curvature that more closely matches the experimental trend. Nevertheless, the experimental curve remains noticeably flatter than both computational predictions, suggesting that the temperature dependence of the NTE coefficient is weaker than captured by either approach.

\begin{figure}
\includegraphics[scale=0.30]{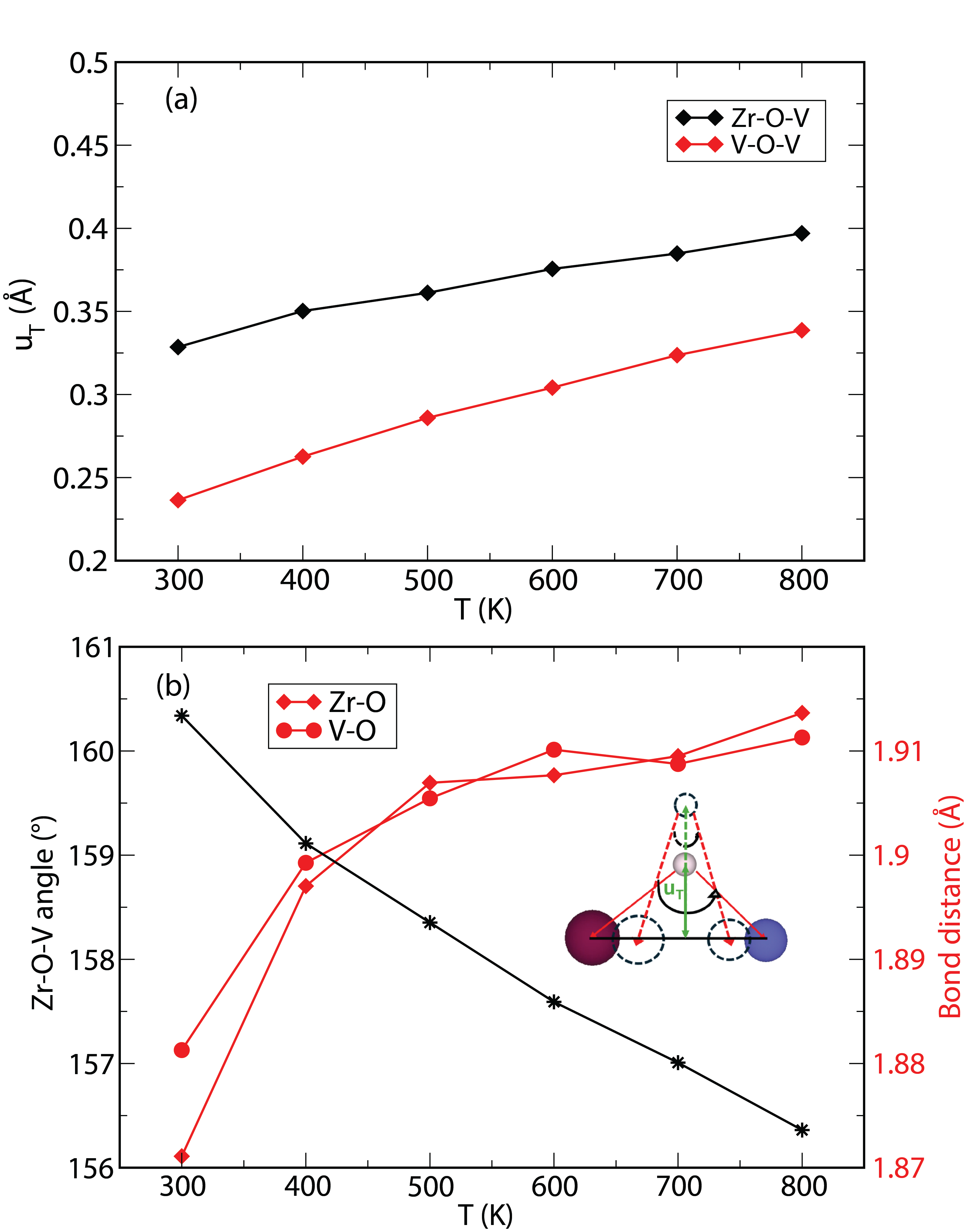} \\
\caption{(a) Comparing the temperature dependence of transversal displacement (u$_T$) for Zr--O--V and V--O--V linkages obtained from ML-MD. (b) Change in Zr--O and V--O bond lengths (right axis) and the Zr--O--V bond angle (left axis) with temperature. The accompanying schematic illustrates how correlated displacements of Zr, V, and O atoms modify local bond lengths and bond angles, thereby governing u$_T$ and the resulting thermal expansion behavior.}
\label{fig:uT}
\end{figure}

Radial distribution functions (RDFs) for Zr--V, Zr--O, and V--O atomic pairs were then analyzed at each temperature to examine changes in the local atomic environments, as shown in Fig. S8. 
The positions of the first RDF peaks, corresponding to the nearest neighbor distances, are located at 2.01 {\AA}, 3.9 {\AA}, and 1.87 {\AA} for Zr--O, Zr--V, and V--O, respectively, in good agreement with the experimentally reported average bond lengths at 300 K.~\cite{Evans_1998}
The RDFs show no significant shift in the Zr--O and V--O nearest-neighbour peaks with increasing temperature, indicating that the most probable bond lengths within ZrO$_6$ octahedra and VO$_4$ tetrahedra remain essentially unchanged. 
However, since RDFs represent a distribution of interatomic distances, subtle changes in individual bond lengths can be masked by peak broadening and overlap. Therefore, direct analysis of the average bond distances reveals a systematic increase in both V--O and Zr--O bond lengths with temperature (see Fig.\ref{fig:uT} (b)), which is not immediately apparent from the RDF peak positions. The thermal disorder causes asymmetric broadening of the nearest-neighbor RDF peaks, with tail shifts more pronounced than shifts in peak maxima. This broadening suggests increased bond lengths and local disorder, particularly around oxygen atoms, which are more clearly captured by bond-angle analysis.

To this end, we analyzed bond-angle distributions averaged over the MD trajectories, as shown in Fig.~\ref{fig:angles}.
Across all panels (a-d), the simulated data show well-defined, largely singular peaks and a systematic temperature dependence. 
For the inter-polyhedral angles, single dominant peaks are observed at $\sim$160$^\circ$ (Zr--O--V) and $\sim$168$^\circ$ (V--O--V) at 300 K, which progressively shift toward lower angles at higher temperatures. These systematic shifts arise from mutual rotations of the connected polyhedra, which reduce the inter-polyhedral angles, thus bringing the connected units closer and driving the overall volume contraction of the framework. In contrast, the intra-polyhedral angles (O--V--O and O--Zr--O) show minimal peak shifts and only slight broadening with temperature, implying that the VO$_4$ tetrahedra and ZrO$_6$ octahedra remain largely undistorted internally. To further elucidate the extent of polyhedral deformation, CSM values were obtained for representative configurations selected from the MD trajectories (every 1000$^{th}$ configuration after equilibration) at each temperature, as shown in Fig.~\ref{fig:angles} (e), (f), (g) and (h). The CSM quantifies how much an observed coordination environment deviates from its corresponding ideal reference geometry. Accordingly, values approaching zero indicate a geometry close to the ideal polyhedron, whereas increasing CSM values reflect progressively larger structural distortions. The CSM distributions for the V-centered polyhedra are concentrated at low values ($<1$), indicating negligible deformation of the VO$_4$ tetrahedra. In contrast, the broader Zr-centered distributions, together with their pronounced shift toward higher CSM values upon heating, indicate comparatively greater, yet still modest, deformation of the ZrO$_6$ octahedra and thus enhanced structural flexibility. This difference in polyhedral rigidity, as shown by the angle analysis and CSM, is consistent with the bonding analysis, which reveals stronger and more covalent V--O bonds than Zr--O interactions. The stronger V--O bonds preserve the VO$_4$ tetrahedra, while weaker Zr--O interactions allow greater flexibility of the ZrO$_6$ octahedra.
This contrast favors collective rotations over intra-polyhedral distortion, reducing the inter-polyhedral angles upon heating and driving NTE.
The CSM distributions for the V--O--V and Zr--O--V angles exhibit similar trends, with the latter being broader and extending to larger values, reflecting the greater displacement of the bridging O atom. This behavior is consistent with the corresponding angle distributions.

RDFs and angular distributions treat interatomic distances and angles separately, failing to describe how atoms deviate from their ideal positions in a combined way. To address this, we calculate the transversal displacement (u$_T$), a more comprehensive descriptor to measure deviations in both bond lengths and bond angles within a single metric.
Fig.~\ref{fig:uT}(a) shows the temperature dependence of u$_T$ calculated for the Zr--O--V and V--O--V linkages, averaged over the MD trajectories. The significant increase of $\sim$40\% in u$_T$ values for both linkages with increasing temperature indicates strong inter-polyhedral bending, responsible for volume contraction. And, the consistently larger u$_T$ values for Zr--O--V linkages compared to V--O--V, re-affirms that greater distortions arise from the collective motion between the ZrO$_6$ octahedra and VO$_4$ tetrahedra.
This is again consistent with the bonding analysis, which supports the picture of rigid polyhedra maintained by strong intra-polyhedral V--O bonds and connected by weaker, more flexible Zr--O linkages. 
The volume contraction is dominated by the QRUMs, accompanied by modest but discernible stretching of the internal bonds, pointing towards a cooperative mechanism. As shown in Fig.~\ref{fig:uT}(b), both the Zr--O and V--O bond distances increase modestly by approximately $\sim$2\% between the lowest and highest temperatures, indicating slight local bond stretching. Simultaneously, the Zr--O--V angle decreases monotonically by about $\sim$2.5\%, reflecting progressive bending of the inter-polyhedral linkages. The decrease in the Zr--O--V angle enhances the transverse component of oxygen displacements (as reflected in increasing u$_T$), while the elongation of bonds provides additional flexibility to sustain these distortions without inducing lattice instability. Although the relative changes are comparable, a percentage change in an angle is not directly equivalent to the same percentage change in bond length. The framework volume is considerably more sensitive to reductions in inter-polyhedral angles, as they alter the network geometry and bring neighboring polyhedra closer together. 
These observations are in perfect agreement with previous reports demonstrating similar mechanisms and QRUMs for other framework materials.~\cite{Pryde1996, miller2009negative} 

\section{Conclusion}\label{concl}

In this study, a comprehensive experimental–theoretical analysis of ZrV$_2$O$_7$ provides atomic-scale insight into its temperature-dependent structural behaviour.

QHA and TDEP identify the low-energy vibrational modes associated with NTE while AIMD, by explicitly incorporating anharmonicity, provides a more complete description of the NTE mechanism. 
Importantly, the low-energy phonon modes with negative Gr\"uneisen parameters identified at the QHA level emerge as a computationally inexpensive descriptor for the initial screening of potential NTE materials. 
The observed contraction at the ML-MD level involves minimal polyhedral deformation: VO$_4$ tetrahedra remain particularly rigid, while ZrO$_6$ octahedra exhibit modest distortions, as confirmed by the CSM analysis. This agrees with previous reports on ZrV$_2$O$_7$~\cite{Evans_1998, Pryde1996} and reinforces the established picture of negative thermal expansion in framework materials, such as ZrW$_2$O$_8$, where collective polyhedral rotations dominate over intra-polyhedral deformation. Consistently, the transverse displacement parameter ($u_T$) captures modest bond-length expansion alongside decreasing inter-polyhedral angles, making it a potential descriptor for subsequent screening stage.

It is also anticipated that more accurate exchange–correlation functionals could improve the thermal expansion description at lower temperatures, although they come at significantly higher computational cost. Future investigations could therefore explore the use of more accurate functionals to further refine the prediction of the thermal expansion observed at lower temperatures.
Overall, this study establishes a consistent microscopic picture connecting lattice dynamics, atomic distortions, and electronic structure to the NTE behavior of ZrV$_2$O$_7$. The combined experimental and first-principles approach identifies the microscopic mechanisms governing high-temperature NTE behaviour, providing suitable descriptors for future research into guided screening of framework materials.

\section{Author contribution}
\textbf{Joana Bustamante}: Methodology (QHA, QHA$_{mod}$, TDEP, and ML-MD), Data curation, Formal analysis, Visualization, Validation, Writing - original draft; \textbf{Aistė Miliūtė}: Investigation (experimental XRD), Methodology (QHA and QHA$_{mod}$), Data curation, Formal analysis, Writing - original draft; \textbf{Namrata Jaykhedkar}: Methodology (MD-ML simulations), Data curation, Formal analysis, Visualization, Validation,  Writing - original draft; \textbf{Aakash Naik} : Investigation (\textsc{LOBSTER} calculations and CSM analysis), Software (coding support); \textbf{Jonas Grandel}: Methodology (MLIPs training); \textbf{Franziska Emmerling}: Project administration, Supervision, Resources$\&$ funding acquisition; \textbf{Tomasz M. Stawski}: Conceptualization, Formal Analysis, Project administration, Supervision, Writing - original draft, Resources $\&$ funding acquisition; \textbf{Janine George}: Conceptualization, Formal Analysis, Project administration, Supervision, Writing - original draft, Resources $\&$ funding acquisition. All authors contributed to  Writing – Review $\&$ Editing.

\begin{acknowledgments}
We thank Florian Matthes for discussions on TDEP methodology and for his assistance with the initial TDEP calculations.
The research was supported by ERC Grant \textsc{MultiBonds} (grant agreement no. 101161771; funded by the European Union. Views and opinions expressed are however those of the author(s) only and do not necessarily reflect those of the European Union or the European Research Council Executive Agency. Neither the European Union nor the granting authority can be held responsible for them.) The research was additionally supported by the Deutsche Forschungsgemeinschaft (DFG) under grant number ZE 1010/15-1 and project number 459785385. Furthermore, we would like to acknowledge the Gauss Centre for Supercomputing e.V. (www.gausscentre.eu) for funding this project by providing generous computing time on the GCS Supercomputer SuperMUC-NG at Leibniz Supercomputing Centre (www.lrz.de) (Project No. pn73da).
We acknowledge DESY (Hamburg, Germany), a member of the Helmholtz Association HGF, for the provision of experimental facilities. Parts of this research were carried out at PETRA III. Data was collected using P02.1: Total Scattering (PDF) operated by DESY Photon Science. We would like to thank Dr. Alba San Jose Mendez for assistance during the experiments. Beamtime was allocated for proposal I-20230110 for AM and TMS. ChatGPT and Microsoft Copilot have been used to improve the clarity, grammar, and readability of selected sections of the manuscript. Copilot was also used to generate potential titles for the manuscript and to provide feedback on earlier versions of the manuscript. All generated titles and abstract suggestions have been carefully reviewed and adapted, and the authors take full responsibility for the final content of the manuscript.
\end{acknowledgments}

\section*{Data Availability Statement}

The raw computational data required to reproduce the results of this study are available through our Zenodo repository https://doi.org/10.5281/zenodo.23339013. Post-processing and plotting scripts used to generate figures presented in this work are openly accessible on our GitHub repository https://github.com/DigiMatChem/paper-ZrV2O7-thermal-expansion-sim. Detailed instructions for reproducing each dataset and figure are provided in the accompanying README file.

\section*{References}

\bibliography{main}
\bibliographystyle{rsc} 

\end{document}


\hyphenpenalty=50000 
\tolerance=1000
\emergencystretch=3em

\maketitle
\section{QHA}

\begin{figure}[!h]
\centering
\includegraphics[scale=0.40]{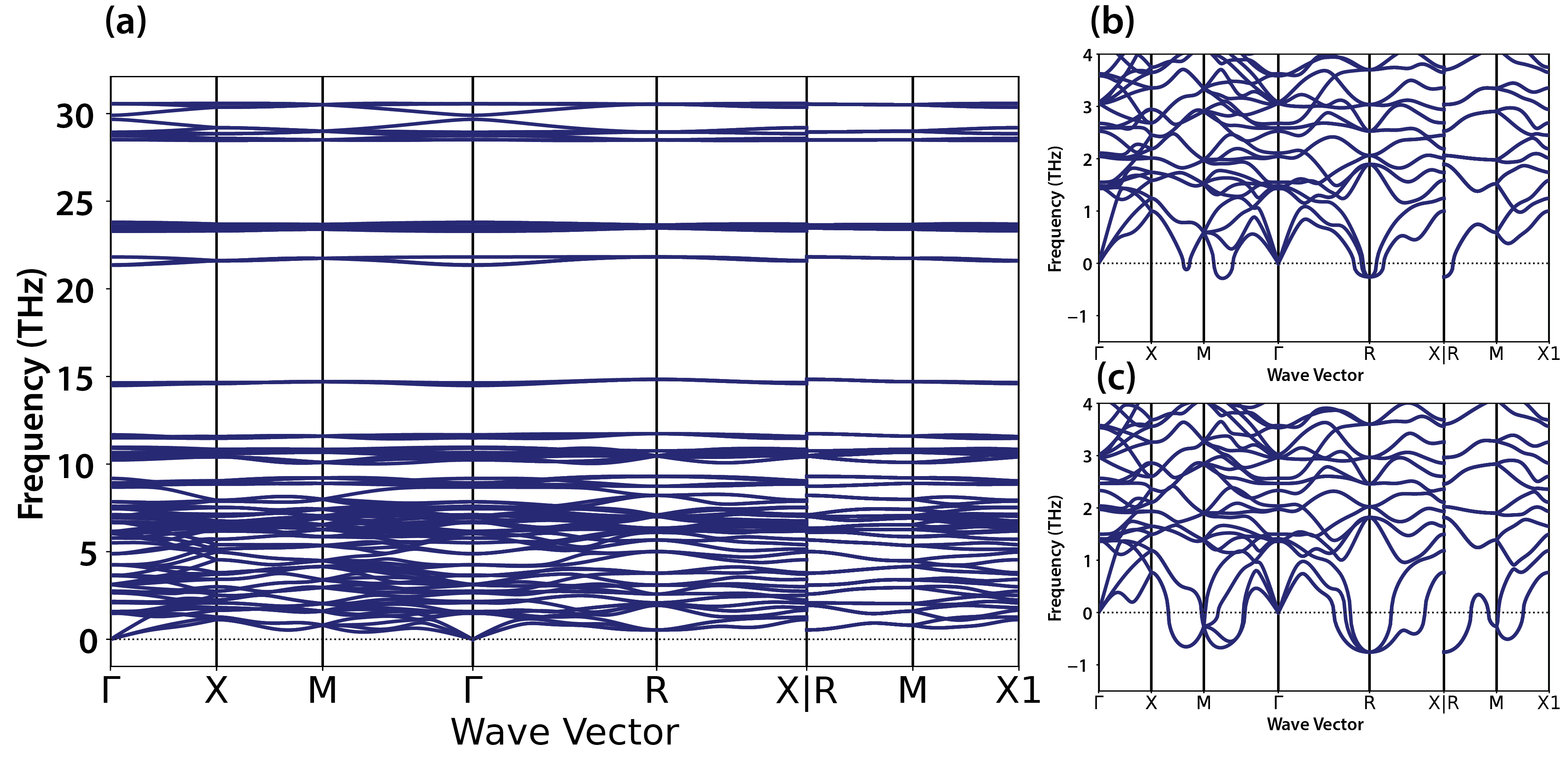}
\caption{Phonon dispersion curves at (a) ground state (V$_0$), (b) -1\%V$_0$, and (c) -3\%V$_0$.}
\label{fig:si_PBE_d3}
\end{figure}

Vibrational properties were calculated at the PBE+D3 level of theory within the finite-displacement method, employing a (2$\times$2$\times$2) supercell as discussed in the main text. Although the equilibrium structure is dynamically stable, imaginary phonon frequencies emerge upon lattice contraction (V < 0.9985$^3$$\times$V$_0$), indicating increasing dynamical instability of the cubic phase with decreasing volume. Such softening under contraction is consistent with the volume sensitivity of the low-frequency modes associated with the NTE behavior. At larger contractions, more pronounced imaginary branches may also signal structural instability and could indicate a phase transition, as commonly observed in framework materials exhibiting NTE. Figure~\ref{fig:si_PBE_d3} illustrates this evolution by comparing the phonon dispersions obtained at the equilibrium volume with those calculated at contractions of -1\% (0.997$^3$$\times$V$_0$) and -3\% (0.99$^3$$\times$V$_0$).

\subsection{Thermal expansion coefficient}\label{sec:si_tec}
The Gr\"uneisen parameter quantifies the relationship between the volume dependence of phonon frequencies and the macroscopic thermal response. Within the quasi-harmonic approximation, the volumetric thermal expansion coefficient ($\alpha_V$) of a cubic crystal can be expressed as:

\begin{equation}
\alpha_V = \frac{\overline{\gamma}C_V}{B\,V_0}
\end{equation}

\begin{figure}[!h]
\centering
\includegraphics[scale=0.55]{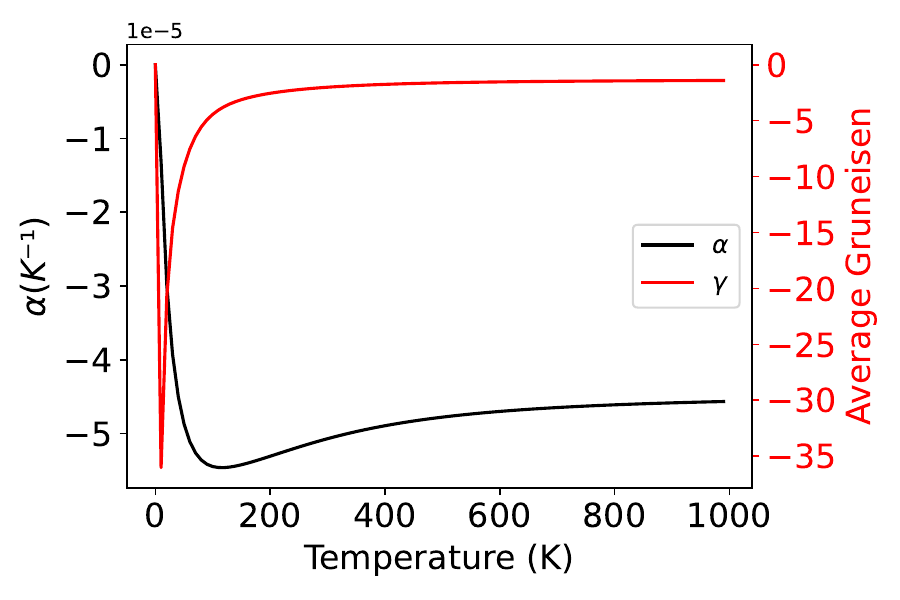}
\caption{Temperature-dependent average Gr\"uneisen parameter ($\overline{\gamma}$) and volumetric thermal expansion coefficient ($\alpha_V$) of cubic ZrV$_2$O$_7$, calculated at the PBE+D3 level within the quasi-harmonic framework. The negative values of both quantities across the investigated temperature range are consistent with the material's negative thermal expansion behavior.}
\label{fig:si_Gru_alpha}
\end{figure}

Where ($\overline{\gamma}$) is the average Gr\"uneisen parameter, ($C_V$) is the heat capacity,  ($B$) is the isothermal bulk modulus, and ($V_0$) is the molar volume. 
Figure~\ref{fig:si_Gru_alpha} illustrates the temperature dependence of the average Gr\"uneisen parameter and the volumetric thermal expansion coefficient. The average Gr\"uneisen parameter remains negative throughout the investigated temperature range, leading to a consistently negative volumetric thermal expansion coefficient. This behavior is consistent with the softening of low-frequency phonon modes under compression and further supports the predicted NTE behavior of ZrV$_2$O$_7$ within the quasi-harmonic framework.

\begin{figure}[!h]
\centering
\includegraphics[scale=0.43]{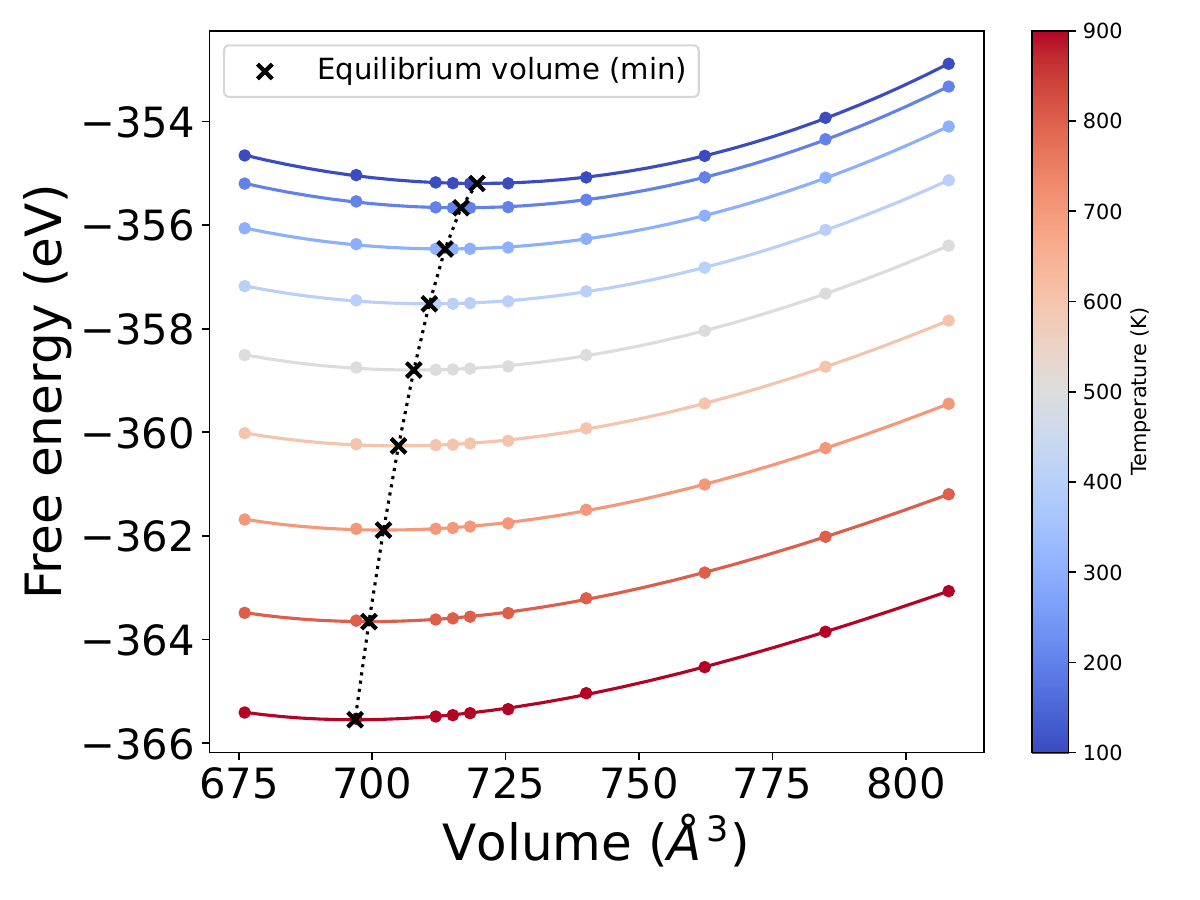}  
\caption{Total free energy as a function of volume for temperatures 100 K to 900 K. Symbols indicate the calculated values, obtained using vibrational free energies at sampled volumes and linear extrapolation for configurations at smaller volumes. Solid lines correspond to fits using the Vinet equation of state. The minima of the fitted curves determine the equilibrium volume at each temperature. }
\label{fig:si_vinet_fitting}
\end{figure}

The temperature-dependent equilibrium volumes were further evaluated using the modified quasi-harmonic approximation (QHA$_{\mathrm{mod}}$) as described in the main text. Because imaginary phonon frequencies emerge at contracted volumes, the corresponding vibrational free energies were estimated by linear extrapolation from dynamically stable configurations. The resulting total Helmholtz free energies were fitted using the Vinet equation of state at each temperature, allowing the equilibrium volumes to be determined from the minima of the fitted curves.
As shown in Figure~\ref{fig:si_vinet_fitting}, the free-energy minima progressively shift toward smaller volumes with increasing temperature, reproducing the characteristic lattice contraction associated with NTE. The agreement between the Gr\"uneisen-based thermal expansion analysis and the temperature-dependent free-energy minimization demonstrates that QHA$_{\mathrm{mod}}$ captures the overall NTE trend, supporting its potential application as a computationally efficient approach for screening candidate NTE materials.
Nevertheless, the emergence of imaginary phonon modes upon compression highlights the limitations of the harmonic description at contracted volumes. Although QHA$_{\mathrm{mod}}$ provides a practical way to estimate thermal expansion behavior, it does not explicitly account for temperature-induced phonon renormalization. This limitation motivates the inclusion of explicit anharmonic effects, as explored through the TDEP calculations discussed in the main text and the following section.

\newpage

\section{TDEP}

To complement our QHA$_{mod}$ analysis and explicitly account for temperature-induced phonon renormalization, vibrational properties of ZrV$_2$O$_7$ were further investigated using the temperature-dependent effective potential (TDEP) approach. In contrast to the QHA and QHA$_{mod}$, where phonon frequencies depend on volume while the lattice dynamics remain harmonic at each fixed volume, TDEP determines effective second-order force constants from thermally representative atomic configurations. The resulting force constants therefore incorporate, in a renormalized manner, the influence of higher-order anharmonic interactions on the finite-temperature phonon spectrum. 
\begin{figure}[!h]
\centering
\includegraphics[scale=0.40]{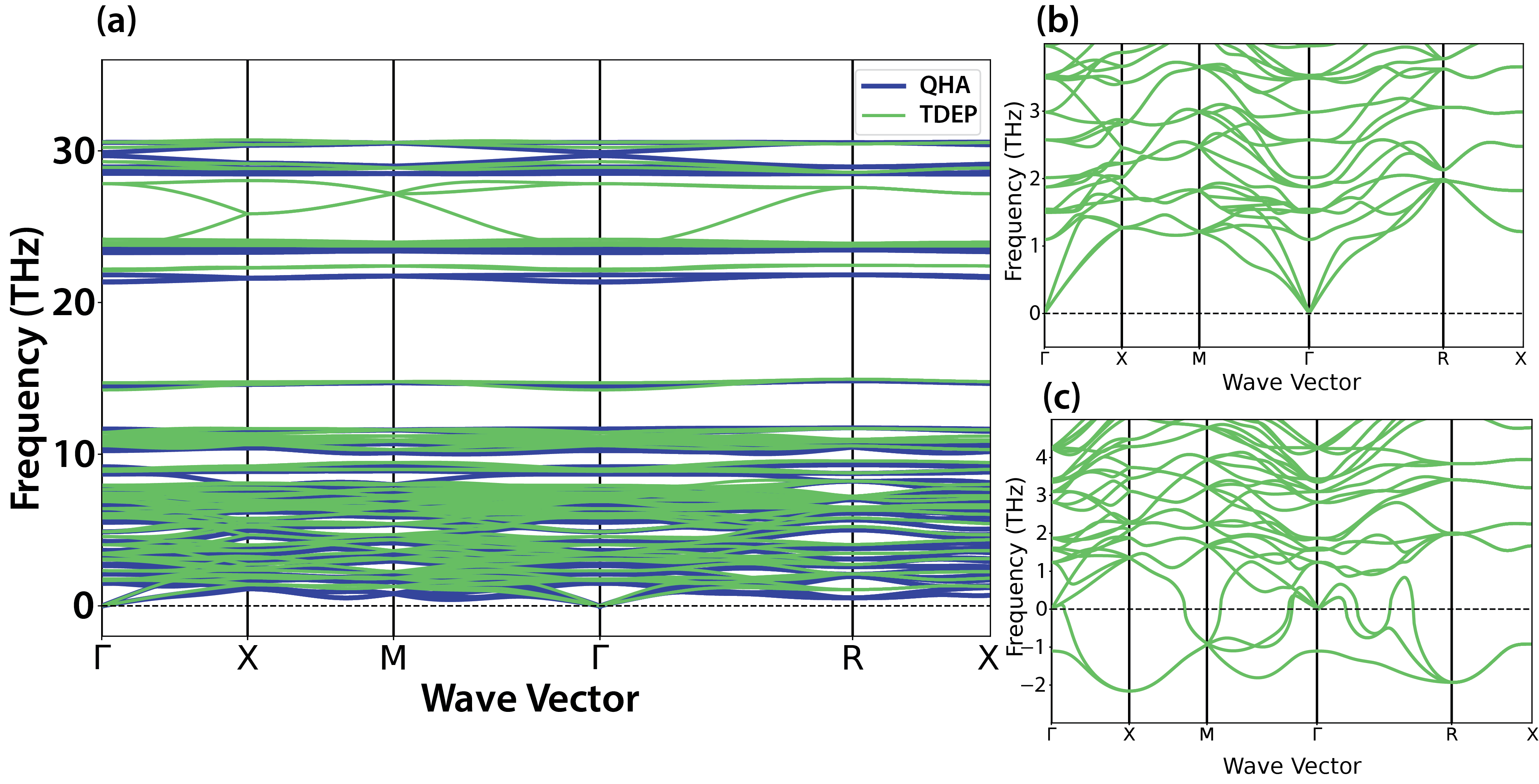}\\
\caption{Phonon dispersion curves comparison for TDEP (300 K) and DFT (0K) at (a) the ground state (V$_{0}$), (b) 0.99$^3$ × V$_{0}$ contraction, and (c) 0.98$^3$ × V$_{0}$.}
\label{fig:si_tdep_comp}
\end{figure}
\begin{figure}[!h]
\centering
\includegraphics[scale=0.45]{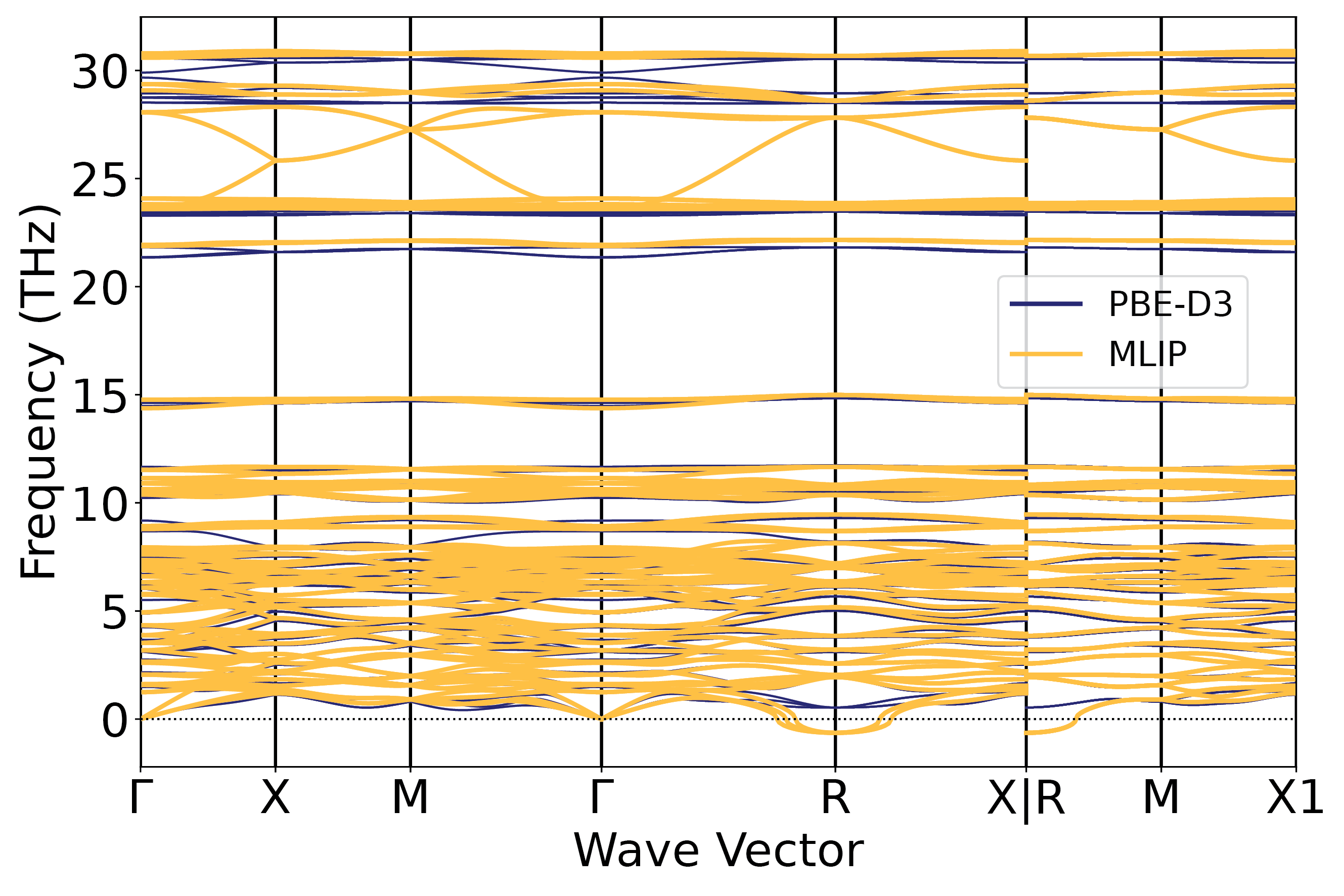} 
\caption{Comparison of phonon dispersion curves using PBE+D3 and MLIP }
\label{fig:si_MLFF_validation_tdep}
\end{figure}

The phonon dispersion obtained from the TDEP-derived effective force constants was compared with the corresponding QHA phonon spectrum Figure~\ref{fig:si_tdep_comp}. At the ground state (at 300 K for TDEP and 0 K for the QHA), both approaches reproduce the main features of the phonon dispersion, including the characteristic separation between the medium- and high-frequency vibrational regions. Nevertheless, the TDEP spectrum shows a slight softening of the higher-frequency optical modes, which we believe has very little impact on the thermal expansion coefficient analysis, as the NTE behavior is predominantly governed by low-frequency modes with negative Gr\"uneisen parameters. Furthermore, the good agreement between the phonon dispersion curves obtained directly from PBE+D3 calculations and those predicted by the machine-learned interatomic potential (MLIP), as shown in Figure~\ref{fig:si_MLFF_validation_tdep}, supports the reliability of the MLIP trained from the PBE+D3 functional. These results confirm that the NTE behaviour arises from the finite-temperature treatment rather than from inaccuracies of the MLIP.

The effect of temperature-induced phonon renormalization becomes particularly evident upon lattice contraction. Within the QHA, imaginary phonon frequencies emerge at contracted volumes, indicating a loss of dynamical stability. At a small contraction of V=0.99$^3$ × $V_0$~\ref{fig:si_tdep_comp} (b)), however, the corresponding TDEP phonon spectrum remains dynamically stable, demonstrating that finite-temperature anharmonic effects can stabilize modes that are unstable within the harmonic approximation. At higher contractions (V<~0.99$^3$ × $V_0$)~\ref{fig:si_tdep_comp}(c)), imaginary frequencies also emerge within the TDEP-derived spectra, indicating that anharmonic renormalization is no longer sufficient to stabilize these modes. The persistence of these instabilities at smaller volumes may indicate structural instability of the cubic phase and could represent a precursor to a structural phase transition under the corresponding thermodynamic conditions. Overall, these results demonstrate that finite-temperature anharmonic renormalization plays a central role in extending the dynamical stability of cubic ZrV$_2$O$_7$ under moderate lattice contraction, while also revealing a regime in which the cubic structure becomes intrinsically unstable.
\begin{figure}[h]
\centering
\includegraphics[scale=0.43]{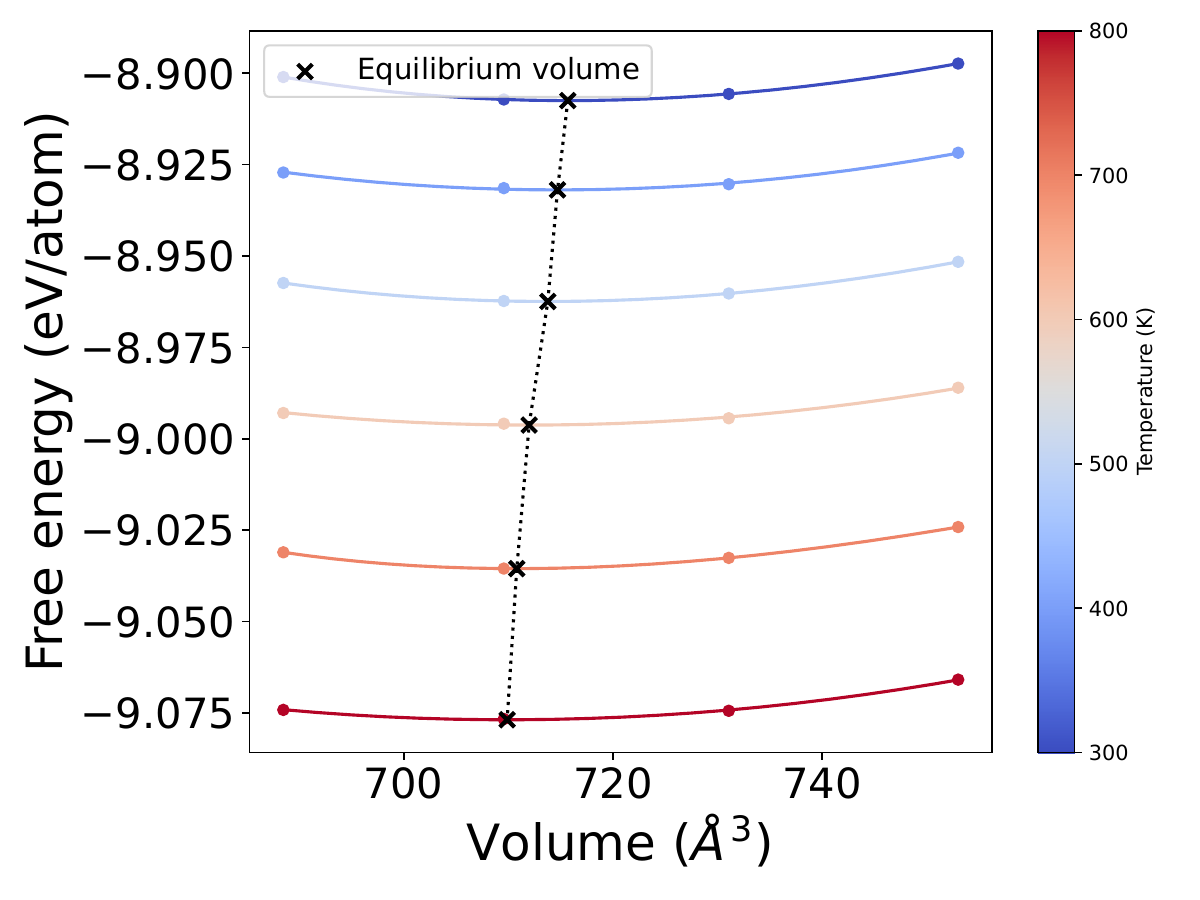} \\
\caption{Free-energy–volume Vinet fitting  for TDEP calculations for cubic ZrV$_2$O$_7$.}
\label{fig:si_Fvib_tdep}
\end{figure}

To determine the temperature-dependent equilibrium volume within the TDEP framework, we evaluated the free energy as a function of volume at each temperature and then fitted it with the Vinet equation of state. The minimum of each fitted free-energy curve defines the equilibrium volume, \($V(T)$\), at the corresponding temperature. Representative Vinet fits over the temperature range from 300 to 800 K are shown in Fig~\ref{fig:si_Fvib_tdep}. As temperature increases, the free-energy minima shift toward smaller volumes, providing direct evidence of the lattice contraction associated with the NTE behavior predicted by TDEP.
The resulting equilibrium volumes were later converted into lattice parameters using the cubic relation $a(T)= V(T) ^{(1/3)}$. The following section compares the temperature-dependent lattice parameters and corresponding thermal expansion coefficients obtained from TDEP with those derived from QHA$_{mod}$, AIMD, and experimental data. This comparison helps clarify the anharmonic effects on the predicted NTE mechanism.

\section{MD}

To validate the accuracy of the trained force fields for the predicted MD trajectories, 40 random structures were sampled and evaluated with single-point DFT to obtain reference energies and forces. We compared these values with the MLIP predictions, which accurately reproduce the DFT energies and forces across the investigated temperature range. At 300 K, the force MAEs are only 0.039-0.041 eV/{\AA}, with corresponding RMSEs of 0.051-0.055 eV/{\AA}, while the energy MAE and RMSE are both 0.000 eV per atom within the reported precision. At 800 K, the force errors increase moderately to 0.069-0.073 eV/{\AA} (MAE) and 0.091-0.096 eV/{\AA} (RMSE), while the energy errors remain very small at 0.000 and 0.001 eV per atom, respectively. The parity plots show a one-to-one correspondence between MLIP and DFT forces, with the data points closely clustered around the ideal agreement line for all three Cartesian components of forces at extreme temperatures. The absence of appreciable systematic deviations or outliers further confirms the ability of the MLIP to accurately reproduce the underlying DFT potential-energy surface.

\begin{figure}[!h]
\centering
\includegraphics[scale=0.29]{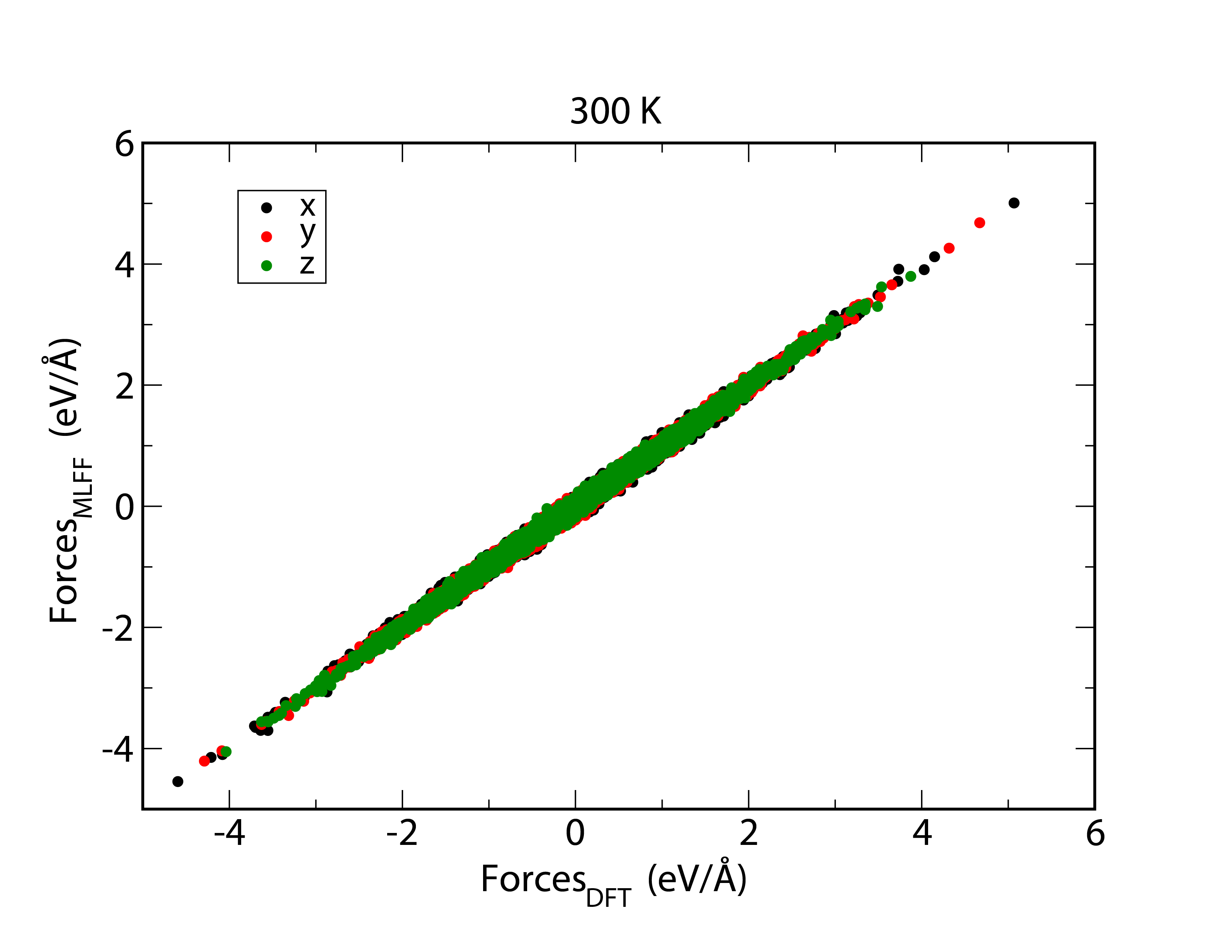} \includegraphics[scale=0.29]{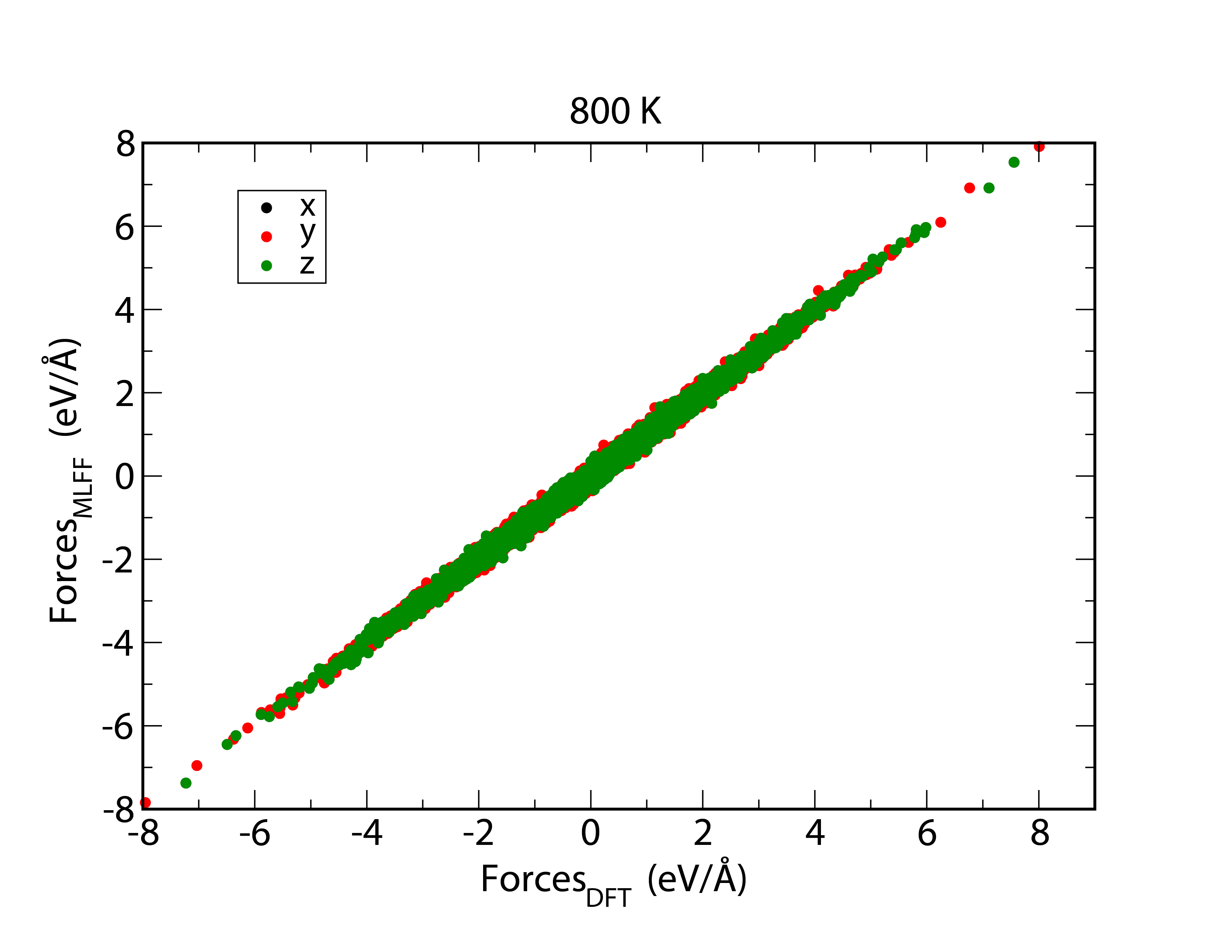} \\
\caption{Comparison of DFT and MLFF predicted atomic force components for ZrV$_2$O$_7$ at low and high temperatures. There is an excellent linear correlation between MLFF and DFT forces at both temperatures, 300 K and 800 K. }
\label{fig:si_MLFF-validation}
\end{figure}
\begin{table}[!h]
\caption{\label{tab:table_s1} Temperature-dependent error metrics, comprising mean error (ME), mean absolute error (MAE), and root mean square error (RMSE) of the machine learning potentials (MLPs) from the MD-production runs for forces and energies.}
\centering
\begin{tabular}{cccccc}
T & Error & \multicolumn{4}{c}{Parameter} \\
\hline
\\
 & & Force (x) & Force (y) & Force (z) & Energy/atom \\
 (K) & & (eV/{\AA}) & (eV/{\AA}) & (eV/{\AA}) & (eV) \\
 \\
 \hline
 \\
300 & MAE & 0.041 & 0.040 & 0.039 & 0.000 \\
& RMSE & 0.055 & 0.054 & 0.051 & 0.000 \\
\\
800 & MAE & 0.069 & 0.073 & 0.069 & 0.000 \\
& RMSE & 0.091 & 0.096 & 0.091 & 0.001 \\
\\
\hline
\\
\end{tabular}
\end{table}

\begin{figure*}[!h]
\includegraphics[scale=0.23]{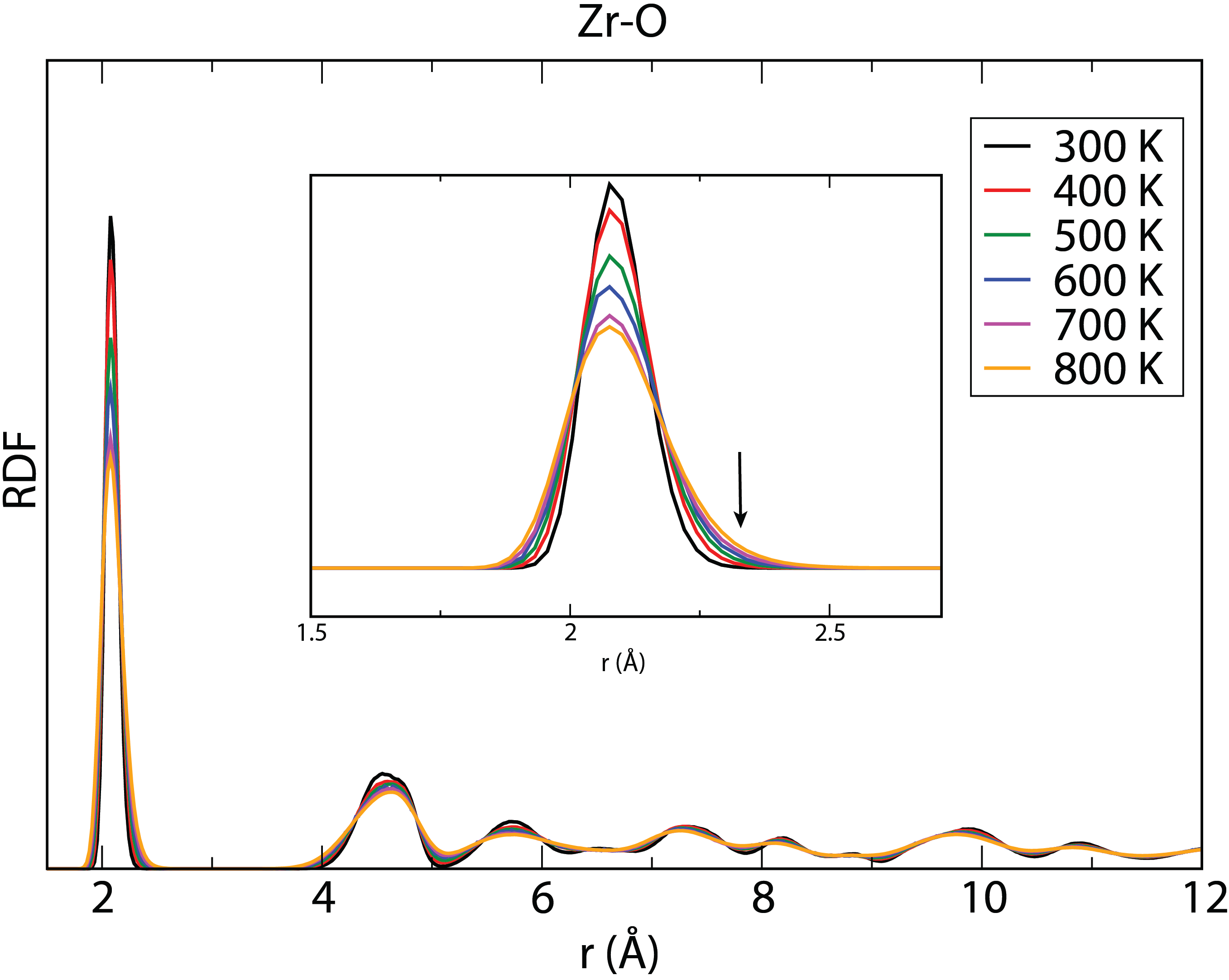} \includegraphics[scale=0.23]{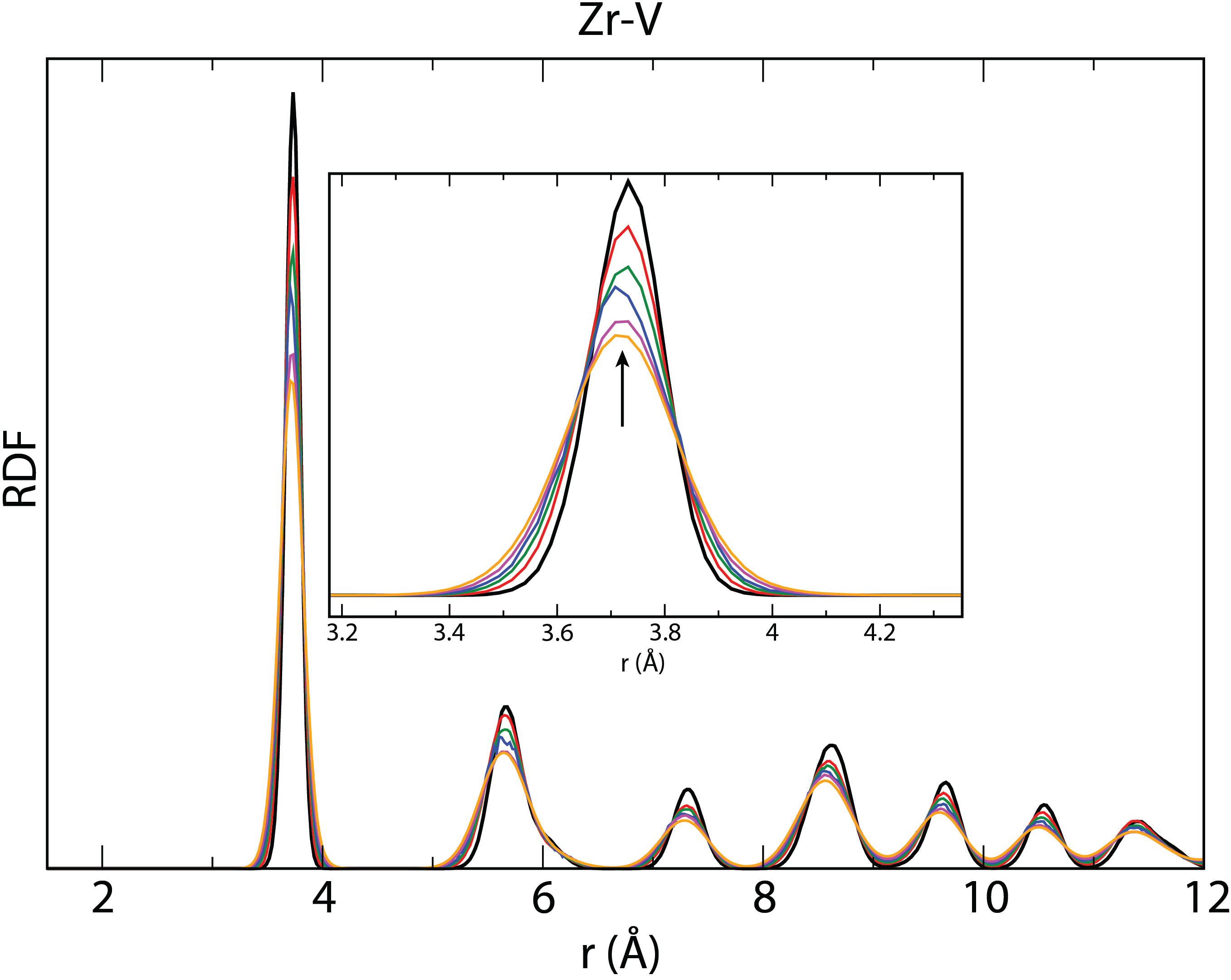} 
\includegraphics[scale=0.23]{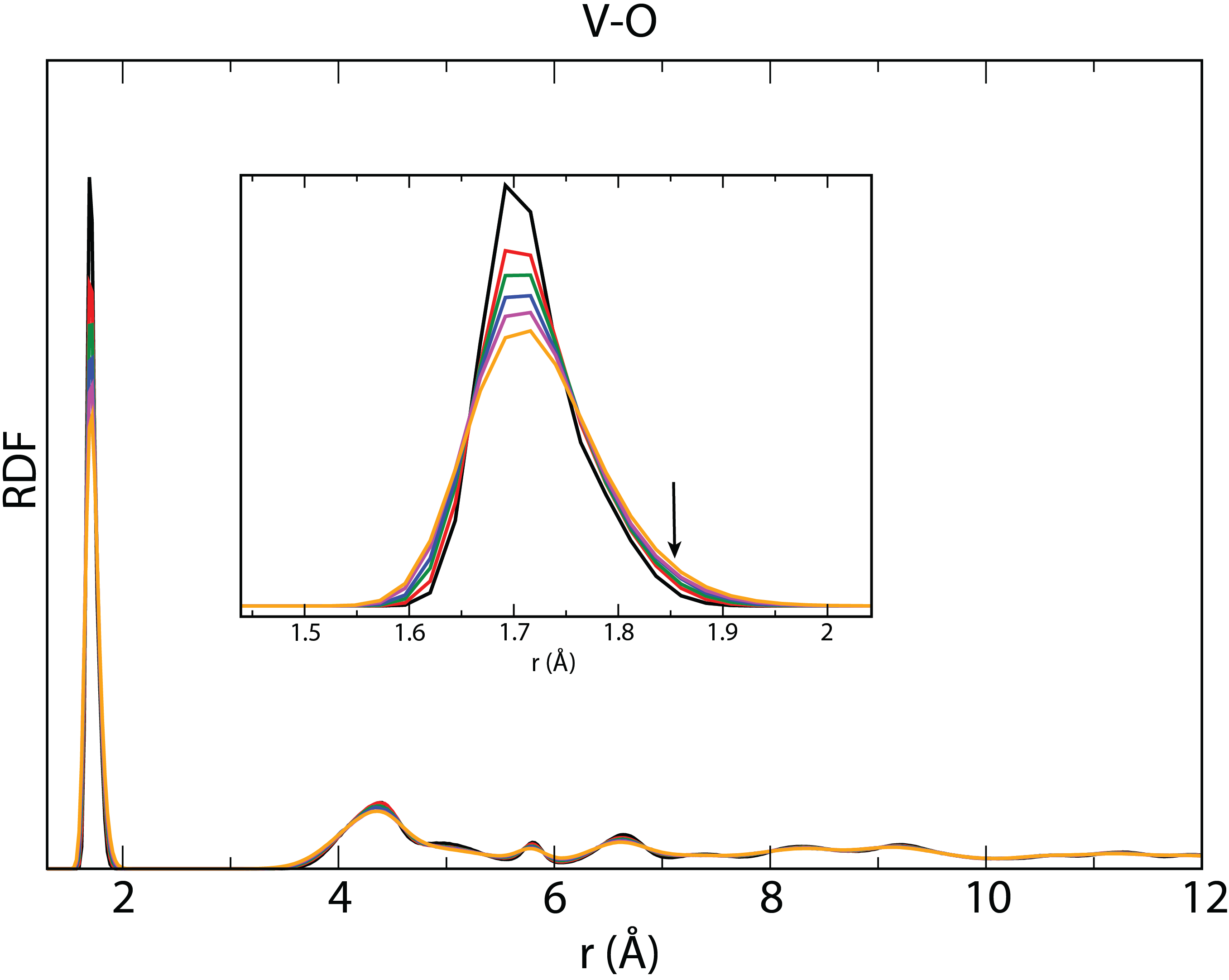} \\
\caption{Radial distribution functions (RDF) for Zr--O, Zr--V, and V--O atomic pairs at various temperatures. A magnified view of the first peak, with marked indications of the peak tails and peak shift, is provided in the insets.}
\label{fig:si_rdf}
\end{figure*}

With increasing $T$, radial distribution functions (RDFs) of all atomic pairs exhibit a gradual broadening of their peaks, consistent with enhanced thermal motion and increased vibrational amplitudes of the atoms. The positions of the first RDF peaks, corresponding to the nearest neighbor distances, are located at 2.01 {\AA}, 3.9 {\AA}, and 1.87 {\AA} for Zr--O, Zr--V, and V--O, respectively, in reasonable or good agreement with the experimentally reported average bond lengths~\cite{}. 
The Zr--V RDF shows relatively sharp, well-defined peaks over a wider radial range, reflecting the persistence of long-range periodicity in the cation sublattice. In contrast, the peaks associated with the Zr--O and V--O correlations become progressively broader with increasing temperature, indicating that oxygen displacements facilitate cooperative distortions or reorientations of the connected ZrO$_6$ octahedra and VO$_4$ tetrahedra.
The RDFs show no significant shift in the Zr--O and V--O nearest-neighbour peaks with increasing temperature, indicating that the most probable bond lengths within ZrO$_6$ octahedra and VO$_4$ tetrahedra remain essentially unchanged. In contrast, the first Zr--V peak shifts slightly toward shorter distances (approximately 0.04 {\AA} between the lowest and highest temperatures), suggesting that changes in cation distances contribute to the macroscopic volume contraction.
However, since RDFs represent a distribution of interatomic distances, subtle changes in individual bond lengths can be masked by peak broadening and overlap. Direct analysis of the average bond distances reveals a systematic increase in both V--O and Zr--O bond lengths with temperature (see Figure 7 (b) in the main text), which is not immediately apparent from the RDF peak positions. The temperature-induced thermal disorder leads to asymmetric broadening of the nearest-neighbour peaks, resulting in only a modest shift of the peak tails rather than a clear displacement of the peak maxima. The observed asymmetry and tailing in the magnified RDF insets therefore indicate a gradual increase in the average bond lengths. These results show that the local structural response is governed by subtle bond-length changes and increasing disorder, particularly involving oxygen atoms, which direct bond-angle analysis can capture better.

\section{Experimental structure}
\begin{figure}[!h]
\centering
\includegraphics[scale=0.4]{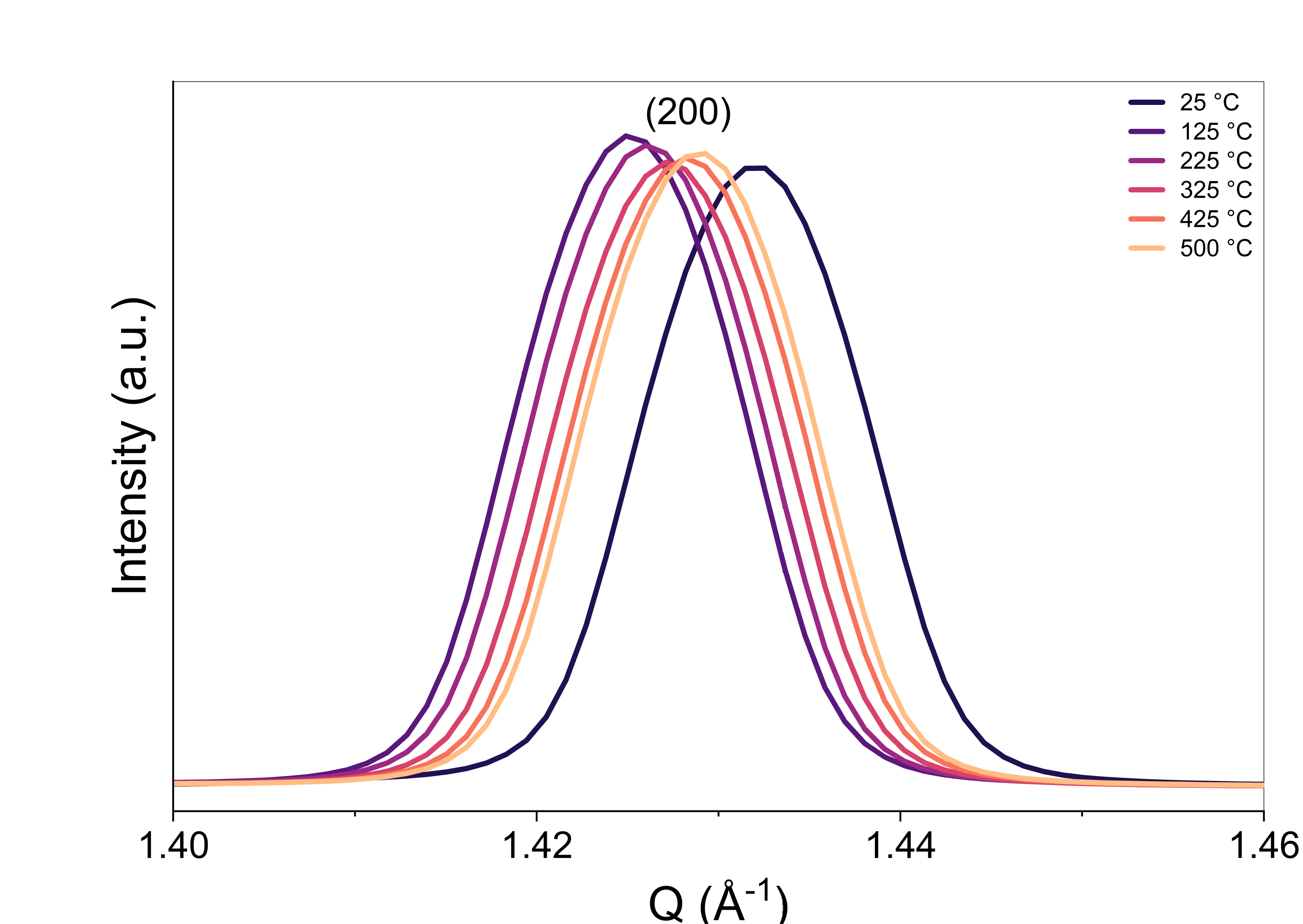} \\
\caption{X-ray diffraction (XRD) (200) reflection of ZrV$_2$O$_7$ showing a peak at 1.4 to 1.46 {\AA}$^{-1}$ shift with temperature. Exact experimental temperatures of 25, 125, 225, 325, 425, and 500 °C correspond to $\sim$300, 400, 500, 600, 700, and 770 K, respectively.}
\label{fig:si_dos}
\end{figure}

The experimental lattice contraction and expansion behavior was extracted from the XRD peak shifts using Bragg’s Law:
$\lambda = 2d \sin\theta$, where $\lambda$ is the fixed X-ray wavelength, $d$ is the spacing between crystal planes, and $2\theta$ is the diffraction angle. In the present experiment, a fixed X-ray wavelength of $\lambda = 0.20735$~\AA\ was used, corresponding to a photon energy of 59.79 keV.
Since the experimental diffraction patterns are presented as a function of the scattering vector ($q$), its relationship to the diffraction angle is given by:

\begin{equation}
q = \frac{4\pi}{\lambda} \sin\left(\frac{2\theta}{2}\right) = \frac{4\pi}{\lambda} \sin(\theta)
\end{equation}

This demonstrates that the diffraction angle (\(2\theta \)) is inversely proportional to the interplanar lattice spacing (\(d\)). As \(\lambda \) remains constant during an experiment, any change in the physical dimensions of the crystal lattice directly alters the angle at which X-rays diffract.
In situ XRD characterization of the ZrV$_2$O$_7$ sample showed the reflections shifting to lower angles (left) from room temperature to 300 K, corresponding to an increase in spacing between lattice planes and therefore lattice expansion. From 300 K onward, the opposite behavior of lattice contraction was recorded, corresponding to a shift towards higher angles (right) and a decrease in lattice planes spacing.